\documentclass[usenatbib]{emulateapj}
\usepackage{subfigure}
\newcommand{\etal}{{\it et~al.\/\ }}

\newcommand{\Msun}{M$_{\sun}$}

\shorttitle{Cold Flows and Galactic Disks} 
\shortauthors{Brooks et al.}

\begin{document}

\title{The Role of Cold Flows in the Assembly of Galaxy Disks}

\author{A.\,M. Brooks\altaffilmark{1,2},
         F. Governato\altaffilmark{3},
         T. Quinn\altaffilmark{3},
         C.\,B. Brook\altaffilmark{4},
 	J. Wadsley\altaffilmark{5} 
}
\altaffiltext{1}{California Institute of Technology, M/C 130-33, Pasadena, CA, 91125}
\altaffiltext{2}{e-mail address: abrooks@tapir.caltech.edu }
\altaffiltext{3}{Astronomy Department, University of Washington, Box 351580, Seattle, WA, 98195-1580}
\altaffiltext{4}{Centre for Astrophysics, University of Central Lancashire, Preston, PR1 2HE, UK}
\altaffiltext{5}{Department of Physics and Astronomy, McMaster University, Hamilton, Ontario, L88 4M1, Canada}


\begin{abstract} 

We use high resolution cosmological hydrodynamical simulations to demonstrate 
that cold flow gas accretion, particularly along filaments, 
modifies the standard picture of gas accretion and cooling onto galaxy 
disks.  In the standard picture, all gas is initially heated to the virial 
temperature of the galaxy as it enters the virial radius.  Low mass 
galaxies are instead dominated by accretion of gas that stays well below 
the virial temperature, and even when a hot halo is able to develop in more 
massive galaxies there exist dense filaments that penetrate inside of the 
virial radius and deliver cold gas to the central galaxy.  For galaxies up to 
$\sim$L$^*$, this cold accretion gas is responsible for the star formation 
in the disk at all times to the present.  Even for galaxies at higher masses, 
cold flows dominate the growth of the disk at early times.  Within 
this modified picture, galaxies are able to accrete a large mass of 
cold gas, with lower initial gas temperatures leading to shorter cooling 
times to reach the disk.  Although star formation in the disk is mitigated 
by supernovae feedback, the short cooling times allow for the growth of stellar 
disks at higher redshifts than predicted by the standard model.    
\end{abstract}




\section{Introduction}
\label{intro}

The classic model for galaxy formation within the cold dark matter (CDM) 
paradigm assumes that gas within galaxy halos initially reaches 
the temperature of the virialized halo.  During the collapse of the halo, 
the process of violent relaxation of the dark matter shocks the gas 
component to the virial temperature.  Any subsequently accreted gas is 
also shocked to the virial temperature as it enters the virial radius 
\citep{rees77,silk77,wr78}. 
This assumption has been routinely adopted by many analytical works 
\citep[e.g.,][]{wf91,kauffmann93,Cole94,AR98,sp99,sb07,ks07,silk07}.
Given a density profile, a cooling radius can be calculated, inside 
of which gas 
can radiate away its energy while conserving its specific angular 
momentum to form a centrifugally supported disk that grows from the inside 
out \citep[e.g.,][]{FE80,dalcanton97,mmw,mao98,vdb00,galics, somerville08}.  
In general, these semi-analytic models (SAMs) have been successful in 
reproducing many of the present day properties of galaxies.

Some studies have called this virial temperature assumption into question, 
however.  Recent theoretical work has found that, in fact, the majority 
of baryons may never reach the virial temperature as they accrete onto 
galaxies \citep{BD03,DB06,K05,ocvirk08,K08,D08}.  \citet[][hereafter K05 
and K08, respectively]{K05,K08} 
examined how gas gets into galaxies using a N-Body + Smoothed Particle 
Hydrodynamics (SPH) simulation of a cosmological volume.  They found 
that galaxies with halo masses below a few 10$^{11}$ \Msun ~are dominated 
by ``cold'' gas accretion.  This cold gas is never shock heated to the 
virial temperature of the galaxy, and is thought to be accreted 
predominantly along filaments in the cosmic web \citep[e.g.,][]{massey07}.  
Further support for this new paradigm has come from \citet[][hereafter OPT08]
{ocvirk08} using a similarly large cosmological volume simulated with an 
adaptive mesh refinement (AMR) code.  

For galaxy halos to shock heat infalling gas, the cooling rate of gas behind 
the shock must be slower than the compression rate of the infalling gas 
\citep{BD03, DB06}.  This requirement leads to two regimes in which gas 
is able to avoid shock heating, and instead cools onto the galaxy on a 
free fall timescale instead.  First, galaxies below a certain critical mass 
are not massive enough to support a stable shock because their dynamical 
time is shorter than the cooling time of the halo gas.  OPT08 find 
that the halo mass where cold and shocked accretion contribute equally is 
in good agreement with the theoretical analysis of \citet[][hereafter DB06]
{DB06}, with a critical mass of 10$^{11.6}$ \Msun.  This critical 
mass is roughly independent of redshift.

Second, even when a shock is present, cold gas accretion can occur 
along filaments that penetrate deep inside the hot halo.  The ability of 
filaments to feed galaxy growth has been noted for some time \citep{katz93,
KW93,katz94,bj94,bond96,shen06, harford08}. 
K05 find that their cold flow gas accretion is largely along filaments.  
These filaments are efficient at funneling gas to the galaxy, and can 
bring material to the galaxy from a much larger radius than predicted 
in the standard spherical accretion model.  The mass contributed by filaments 
is a strong function of redshift (DB06, OPT08).  

Note that in the first 
case above, when a galaxy is in the low mass regime incapable of supporting 
a stable shock, cold gas accretion can occur both in a quasi-spherical 
fashion and along filaments, while in the second, high mass regime in 
which a hot halo exists, cold gas accretion occurs primarily along preferred 
directions in filaments.

\begin{deluxetable*}{lccccccccr}
\tablecaption{Simulated Galaxy Properties \label{simsum} }
\tablewidth{0pt}
\tablehead{\colhead{simulation} & 
\colhead{M$_{vir}$} & \colhead {R$_{vir}$} & 
\colhead{V$_{circ}$} &  \colhead{T$_{vir}$} & \colhead{$\lambda$} & 
\colhead{z$_{LMM}$} &  \colhead{$\epsilon$} & \colhead{N within R$_{vir}$} \\
 & \colhead{\Msun} & \colhead {kpc} & \colhead{km/s} & \colhead{K} &  &  &  
\colhead{kpc} & \colhead{dm+star+gas} }  
\tabletypesize{\footnotesize}
\startdata
H579	& 3.4$\times$10$^{10}$ & 85  & 40  & 5.1$\times$10$^4$ & 0.03 & $>$5 & 0.17 & $\sim$9.5$\times$10$^5$ \\
DWF1	& 1.4$\times$10$^{11}$ & 135 & 70  & 1.3$\times$10$^5$ & 0.01 & 1.5 & 0.15 & $\sim$5.3$\times$10$^6$ \\
MW1	& 1.1$\times$10$^{12}$ & 270 & 130 & 5.1$\times$10$^5$ & 0.05 & 2.5 & 0.3 & $\sim$4.8$\times$10$^6$ \\
GAL1	& 3.3$\times$10$^{12}$ & 385 & 190 & 1.1$\times$10$^6$ & 0.02 & 2.8 & 0.3 & $\sim$3.7$\times$10$^6$ \\
H277	& 7.1$\times$10$^{11}$ & 230 & 115 & 3.8$\times$10$^5$ & 0.03 & 2.5 & 0.35 & $\sim$2.3$\times$10$^6$ \\
MW1.lo  & 1.1$\times$10$^{12}$ & 265 & 130 & 5.0$\times$10$^5$ & 0.07 & 2.5 & 0.6 & $\sim$5.8$\times$10$^5$ \\
MW1.ad  & 1.1$\times$10$^{12}$ & 270 & 130 & 5.1$\times$10$^5$ & 0.07 & 2.5 & 0.6 & $\sim$3.0$\times$10$^5$ \\
\enddata
\end{deluxetable*}

Galaxies above the critical mass for shocking appear to be dominated by hot
gas accretion as in the standard model, but they may have been fed by cold 
flow filaments even after a shock develops, particularly at high $z$, and 
they are also built in part by the mergers of smaller galaxies that were 
dominated by cold flows.  Thus, cold flow gas accretion is an 
important component of galaxy formation that has remained largely 
ignored until recently.

Previous work on cold flow gas accretion in simulations have addressed the 
issue using large cosmological volumes, furnishing a statistical description 
of how galaxies get their gas.  This provides a useful starting point for 
modifying semi-analytic models of galaxy formation.  However, with large 
volumes and many simulated galaxies, numerical resolution for any individual 
galaxy must 
be sacrificed.  Hence, these simulations were unable to resolve the internal 
structure of galaxies and investigate the impact of cold flows on the 
formation of galactic {\it disks}.  OPT08 follow gas as it enters the virial 
radius of galaxies, and examined the temperature history of that gas as it 
flows toward the central galaxy, down to 0.1 R$_{vir}$.  However, limited 
resolution prevented them, or K05 or K08, from being able to resolve disks in 
their galaxies.  Thus, the impact of cold flows to disk assembly, and in 
particular the growth of the stellar disk component, has yet to be addressed. 

Rather than sample a  cosmological volume at low resolution, in this 
paper we investigate gas accretion in several very high resolution 
simulations, in order to study the impact of cold gas accretion on the 
assembly of stellar disks.  Each of these galaxies has at least a million 
particles within the virial radius by redshift zero.  This high resolution 
provides two advantages over previous simulation work on this subject.  
First, it allows us to resolve the internal structure of our galaxies and 
their disks.  Second, we overcome artificial broadening of shocks in SPH 
so that we are able to identify entropy increases associated with accretion 
shocks, allowing us to explicitly search for shocks rather than simply 
using a temperature threshold as in previous works (K05, K08, OPT08).  

We are thus capable of examining the temperature and entropy 
history of accreted gas that
eventually settles to the disk and forms stars.  These simulated 
galaxies are run within a $\Lambda$CDM context and span two orders of 
magnitude in galaxy halo mass, allowing us to study the role of cold 
flows in galaxies from 0.01 L$^*$ to L$^*$.  

This paper is outlined as follows: In Section \ref{sims} we briefly discuss 
the simulations used in this study.  In Section \ref{gasaccr} we describe our 
method to identify shocked and unshocked gas accretion to galaxies, 
and find that the fraction of accreted gas that is shocked is a strong 
function of galaxy mass, in agreement with previous works.  In 
Section \ref{diskgrowth} we extend the results of Section \ref{gasaccr}
to investigate the role that unshocked, cold gas accretion 
has in building the stellar disk of 
galaxies, as a function of galaxy mass.  We briefly compare our results 
on disk growth to analytic models of disk galaxy formation.  In Section 
\ref{issues} we discuss the various effects of our numerical scheme on 
the results, and conclude that our results are robust (those readers 
interested strictly in the science results may wish to skip this section).  
Finally, in Section \ref{conclusions} we discuss the implications of 
our results for existing analytic models of disk galaxy formation, and 
conclude.

\section{The Simulations}
\label{sims}

The simulations used in this study are the culmination of an effort to 
create realistic disk galaxies.  As discussed in \citet{G07}, at $z$ = 0 
disk galaxies in our simulations overcome the angular momentum catastrophe 
and resolve the internal structure of disks \citep[see also][which shows 
that galaxies in this paper lie on the Tully-Fisher relation]{G08}.  This 
crucial improvement allows us to undertake the current investigation, and 
is made possible both by dramatically increased numerical resolution and 
by the inclusion of a simple but physically motivated recipe to describe 
star formation (SF) and the effects of subsequent supernova (SN) feedback
\citep[full details in][]{Stinson06}.  Our adopted SN feedback and cosmic 
UV background \citep{HM96} drastically reduce the number of galaxy 
satellites containing a significant stellar population, avoiding the well 
known ``substructure problem'' \citep{wf91, kauffmann93, Quinn96, Moore99}.  
Our SN feedback is also effective at regulating SF, preventing gas from 
artificially cooling too early and forming stars, and hence we produce 
smaller bulges and halos in subsequent mergers \citep{brook04b}.  
This regulation by SN also 
reproduces the observed stellar mass-metallicity relationship for galaxies, 
both at the present time and at high redshift \citep{Brooks07,maiolino08}.  
At $z$ = 3, these simulations reproduce the metallicities, incidence rate, 
and column density distributions of observed damped Lyman alpha absorbers 
\citep[DLAs][]{pontzen08}.

Four simulated disk ``field'' galaxies were chosen to cover two orders 
of magnitude in halo virial mass, from 3.4$\times$10$^{10}$ \Msun ~to 
3.3$\times$10$^{12}$ \Msun.  The three most massive galaxies have been 
discussed in detail in \citet[][i.e., DWF1, MW1, GAL1]{G07,G08}, but they have 
been rerun at eight times the mass resolution and twice the spatial 
resolution for this study. 
All galaxies are simulated using {\sc Gasoline}
\citep{gasoline} in a flat, $\Lambda$-dominated cosmology.  The three 
most massive galaxies were run using a WMAP year 1 cosmology, 
while a WMAP year 3 cosmology was adopted for the newest, lowest mass 
galaxy (hereafter H579). 
The three least massive galaxies were originally drawn from uniform volumes 
ranging from 25 Mpc to 50 Mpc, while 
GAL1 was selected from a 100 Mpc box.  These galaxies were then resimulated 
at higher resolution using the volume renormalization technique \citep{KW93}, 
which allows us to resolve fine structure while capturing the effect of 
large scale torques.  We adopt the same SN and SF efficiency parameters as 
in \citet{G07} and a Kroupa initial mass function \citep{Kroupa}.

Characteristics of each galaxy at $z$ = 0 are given in Table 1. 
Galaxies and their parent halos were identified using 
AHF\footnote{{\bf{A}}MIGA's {\bf{H}}alo {\bf{F}}inder, available for 
download at http://www.aip.de/People/aknebe/AMIGA} \citep{Knebe, GK04}. 
AHF determines the virial radius, R$_{vir}$, for each halo at each output 
time step using the overdensity criterion for a flat universe following 
\citet{Gross97}.  AHF identifies all halos that exist in the high resolution 
region of our simulations at each output time step, 
and the individual particles that belong to them.  We can thus track 
a single gas particle through the entire simulation, as it is accreted to 
a galaxy, as that galaxy later merges with another, and as the gas particle 
forms stars.  

AHF allows us to identify the main galaxy at each output time step from 
$z$ = 0 to higher $z$, and the particles that belong to it.  In this paper, 
we identify the progenitor halos of the main galaxy all the way back to 
$z$ = 6.  Thus, as can be seen in Table 1, particles are traced back to a 
higher $z$ than the last major merger for 
these galaxies.  For example, our MW1 galaxy undergoes a last major 
merger at $z \sim$2.5, and is easily identified at lower redshifts.  At  
higher $z$, we follow the most massive of the two progenitors 
to study the galaxy's history.  

\citet{G07} and \citet{reed03} showed that the mass function in their 
simulations converged at a minimum of 64 dark matter (DM) particles per 
halo. 
Thus, we require a minimum of 64 DM particles for halo identification 
in AHF.  Halos with fewer particles are considered as ``smooth'' gas accretion 
instead, but in practice these halos are ``dark'' and have no bound baryons 
due to heating from our UV background.  The UV background turns on at $z$ = 6 
and by $z$ = 4 has a significant effect on the lowest mass halos, heating 
their gas so that only halos with $\sim$100 or more particles are of 
sufficient mass to retain bound gas particles (with velocities below the 
escape velocity of the halo).  Therefore,
unresolved halos with fewer than 64 DM particles should contribute little 
or no bound gas, and have no significant impact on the gas accretion study 
in this paper.  Resolution effects are explored further in Section~\ref{uv}.

Our highest (lowest) resolution runs have particle masses of 
4.5$\times$10$^4$ M$_{\odot}$ (2.9$\times$10$^6$ M$_{\odot}$), 
1.6$\times$10$^4$ M$_{\odot}$ (4.3$\times$10$^5$ M$_{\odot}$), and 
4.7$\times$10$^3$ M$_{\odot}$ (1.3$\times$10$^5$ M$_{\odot}$) for dark 
matter, gas, and stars, respectively, and a force resolution of 0.15-0.3 
kpc.  These galaxies were simulated to have a similar dynamical range and 
similar number of particles within the virial radius.  For a more 
detailed discussion of the simulations, see \citet{G07}.

We examine the temperature, entropy, and accretion histories for gas 
particles accreted to the main galaxy in these four simulations.  For all 
cases, each gas particle's properties are determined at $z$ = 6, $z$ = 5, 
and $z$ = 4, after which the values are sampled approximately every 
320 Myr until $z$ = 0.  The gas in these simulations is allowed to cool 
via atomic cooling, adopting the cooling function for a primordial 
(metal free) gas.  We show in Section~\ref{mc} that adding metal line cooling 
to the simulations has no effect on the results presented in this paper.

Major mergers have been defined as having a $\sim$3:1 mass ratio, or lower. 
The galaxies in this sample tend to have relatively quiescent merging 
histories, with the last major merger redshifts being greater than 1.5.  
This similarity is by design, in order to compare gas accretion histories of 
galaxies with similar pasts.  In a future paper, we will explore gas 
accretion and disk assembly in a set of L$^*$ galaxies ($\sim$10$^{12}$ at 
$z$ =0) that span a broader range in space in spin parameters and in last 
major merger redshifts.  

In addition to the four main galaxies studied here, a few other simulations 
have been run to investigate numerical convergence.  MW1 is explored at half 
of the dynamical resolution \citep[MW1.lo, the same resolution simulation 
discussed at length in][]{G07}. This lower resolution run is also studied 
with cooling turned off (the adiabatic case, MW1.ad) and with metal line 
cooling added (MW1.mc).  An additional L$^*$ galaxy, H277, with M$_{vir}$ = 
7.1$\times$10$^{11}$ \Msun, is used to investigate the effects of 
output time step resolution on our temperature and entropy criteria.

\section{Gas Accretion to the Galaxy Halo}
\label{gasaccr}

Before investigating the role of shocked versus unshocked gas in our 
simulations, we first want to identify the gas that has been accreted 
smoothly to our galaxy halos, as opposed to gas that has been accreted 
from other galaxy halos.  It is this smoothly accreted gas that 
will be available to potentially shock as in the classic model. 

\subsection{Smooth vs Clumpy Accretion}
\label{smvcl}

We first find all gas particles that have ever been identified as part 
of the main galaxy, back to $z$ = 6.  AHF allows us to identify those gas 
particles which have been part of other galaxies before merging with the
main galaxy.  Those gas particles that are already in the main halo 
progenitor at $z$ = 6 (i.e., the earliest step at which gas particles are
traced) are labeled as ``early'' accretion.  We adopt an extreme definition 
and label as ``clumpy'' accretion any particles that have at any output step 
belonged to a galaxy halo other than the main galaxy we're considering.  
The majority of this gas will be accreted in merger events (note that 
in this definition, even those particles that come from the smaller 
progenitor in a last major merger get labeled as clumpy; it includes both 
the hot and cold gas accreted in this merger), though a small fraction of 
this clumpy gas may also include gas that was tidally stripped or blown 
out of a galaxy halo and later accreted to the main galaxy we're considering.
The remaining gas that does not meet our ``clumpy'' criterion is labeled 
as ``smoothly'' accreted gas.

\begin{figure*}
\plotone{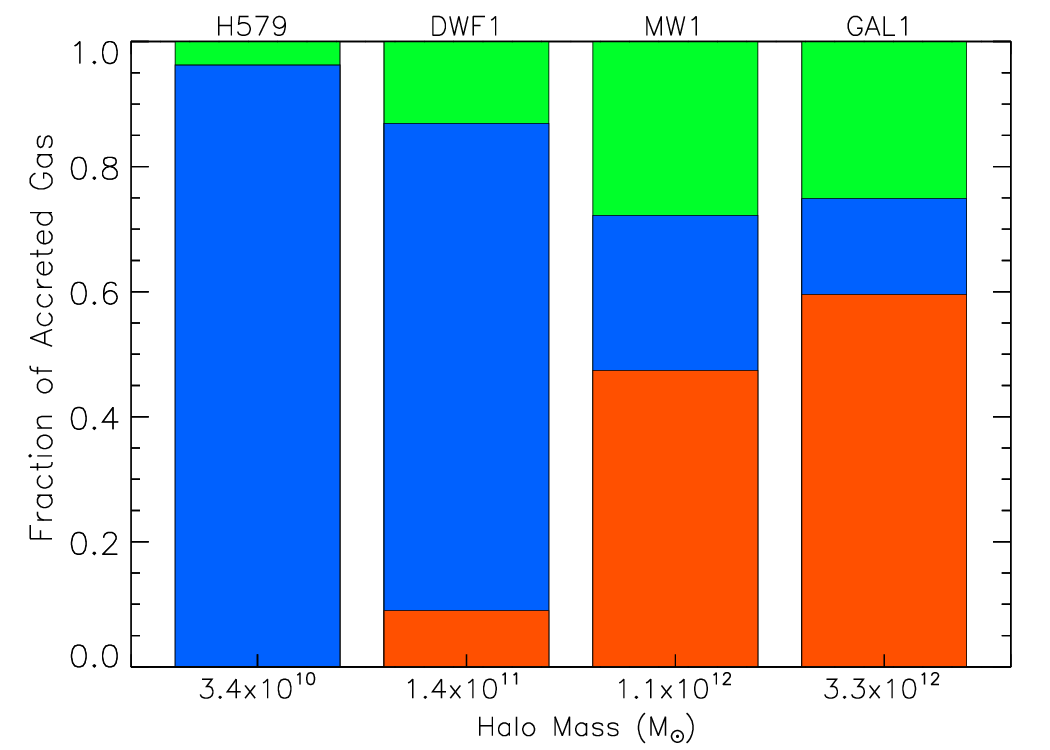}
\caption{
This plot shows the fraction of gas that has been accreted at the 
virial radius since $z$ = 6 that has been accreted as ``clumpy'' (green, 
top), ``unshocked'' (blue), and ``shocked'' (red) gas.  Together, the 
unshocked and shocked gas make up the total of smoothly accreted gas 
that never belonged to another galaxy halo (see Section~\ref{decomp}).  
The total halo masses (in \Msun) of each of the four galaxies are listed 
below their respective bar.  }
\label{fig1} 
\end{figure*}

Fig.~\ref{fig1} shows the fraction of gas accreted since $z$ = 6 that 
has been identified as ``clumpy'' (green, at top of bars) or ``smooth'' 
(the blue and red bars, combined; this separation of smooth is discussed 
in detail in section~\ref{decomp}), in order of increasing galaxy mass for 
our four main galaxies.  The total halo mass (DM+stars+gas) of each of the 
four galaxies is listed below their respective bar.  

The key feature of Fig.~\ref{fig1} is the dominance of smoothly accreted 
gas in the formation of these galaxies, despite our extreme definition
for ``clumpy'' accretion.  The first two columns of Table~\ref{accr} 
list the ratios of ``smooth'' and ``clumpy'' gas accretion to total 
gas accreted to the galaxy since $z$ = 6.  
In all cases, $\sim$75\% or more of the gas accreted to the 
galaxy since $z$ = 6 has never belonged to another galaxy halo (the 
adiabatic case is discussed in section~\ref{adiabatic}).  

Because we resolve halos down to the mass limit where the UV background 
suppresses galaxy formation, the major numerical uncertainty in our 
smooth gas accretion fractions is due to our finite output time interval 
(320 Myr).  Due to output time step resolution, our fractions in 
Table~\ref{accr} for smoothly accreted gas will be an upper limit. 
The issue of output step resolution is explored below (section \ref{timeres}) 
and it is shown that, for a Milky Way type galaxy, increasing time resolution 
will identify a higher fraction of ``clumpy'' material, but not by more 
than 10\%, so smoothly accreted gas still dominates the formation 
of this galaxy. 

The dominance of smooth gas accretion in the baryonic growth of galaxies 
agrees with previous simulation results \citep{K05, 
Murali02}.  \citet{Murali02} looked at this question, in particular.  
Their mass resolution allowed them to fully examine gas accretion in 
galaxies with masses greater than 5.4$\times$10$^{10}$ \Msun ~in halo 
mass, and they found that galaxies above this mass predominantly gain 
mass via smooth accretion rather than in mergers.  In fact, they 
find that mergers account for no more than 25\% of mass accretion at 
redshifts greater than 1, and no more than 35\% of growth at $z$ $<$ 0.5.  
K05 and K08 also concluded that, in a global sense, the growth of galaxies 
is dominated by smooth gas accretion rather than mergers.

\begin{deluxetable}{lcccc}
\tablecaption{Gas Accretion Properties \label{accr}}
\tablewidth{0pt}
\tablehead{
\colhead{galaxy} & \colhead{f$_{\mathrm{smooth}}$} & \colhead {f$_{\mathrm{clumpy}}$} & 
\colhead{f$_{\mathrm{shock}}$} &  \colhead{f$_{\mathrm{unshock}}$} } 
\tabletypesize{\footnotesize}
\startdata
H579	& 0.96 & 0.04 & 0.00 & 1.00 \\
DWF1	& 0.87 & 0.13 & 0.10 & 0.90 \\
MW1	& 0.73 & 0.27 & 0.65 & 0.35 \\
GAL1	& 0.75 & 0.25 & 0.79 & 0.21 \\
H277	& 0.73 & 0.27 & 0.44 & 0.56 \\
MW1.lo  & 0.78 & 0.22 & 0.62 & 0.38 \\ 
MW1.ad  & 0.49 & 0.50 & 0.84 & 0.16 \\ 
\enddata
\tablecomments{The first two columns, f$_{\mathrm{smooth}}$ and 
f$_{\mathrm{clumpy}}$, are the fractional quantities with respect to 
all of the gas accreted to the galaxy since $z$ = 6.  By definition, 
f$_{\mathrm{smooth}}$ and f$_{\mathrm{clumpy}}$ sum to 1.0.  On the 
other hand, the last two columns, f$_{\mathrm{shock}}$ and 
f$_{\mathrm{unshock}}$, are the fractional quantities with respect to 
all of the gas ever {\it smoothly} accreted to the galaxy, and sum to 1.0. }
\end{deluxetable}

Fig.~\ref{fig1} shows that, in general, there is a trend in the amount 
of ``clumpy'' accretion as a function of galaxy mass.  This figure 
suggests that as galaxy masses get smaller, the amount of material accreted 
from other galaxy halos decreases.  This is at odds with the merger trees 
of pure dark matter halos of different masses, which are statistically 
indistinguishable \citep[e.g.,][]{guo08, stewart08, fakhouri08}.
However, Fig.~\ref{fig1} represents only accreted gas, 
and in practice there is a decrease in the baryon fraction of halos with 
decreasing halo mass.  The reionization of the universe acts to unbind 
baryons from halos with V$_{circ}$ below $\sim$30 km/s, and the fraction 
of unbound baryons increases as halo mass decreases further.  Our UV 
background mimics this effect of reionization.  Gas loss 
due to SNe feedback also becomes increasingly effective at these lower 
masses.  The combined effect is for halos to become increasingly 
dark matter dominated toward lower halo mass \citep[e.g.,][]
{bullock00, hoeft06,G07,Brooks07, okamoto08}.  As smaller 
halos merge with even smaller, more dark matter dominated halos, 
the amount of baryonic material accreted from other halos decreases 
with smaller halo masses, as seen in Fig.~\ref{fig1}.

\begin{figure*}
\plotone{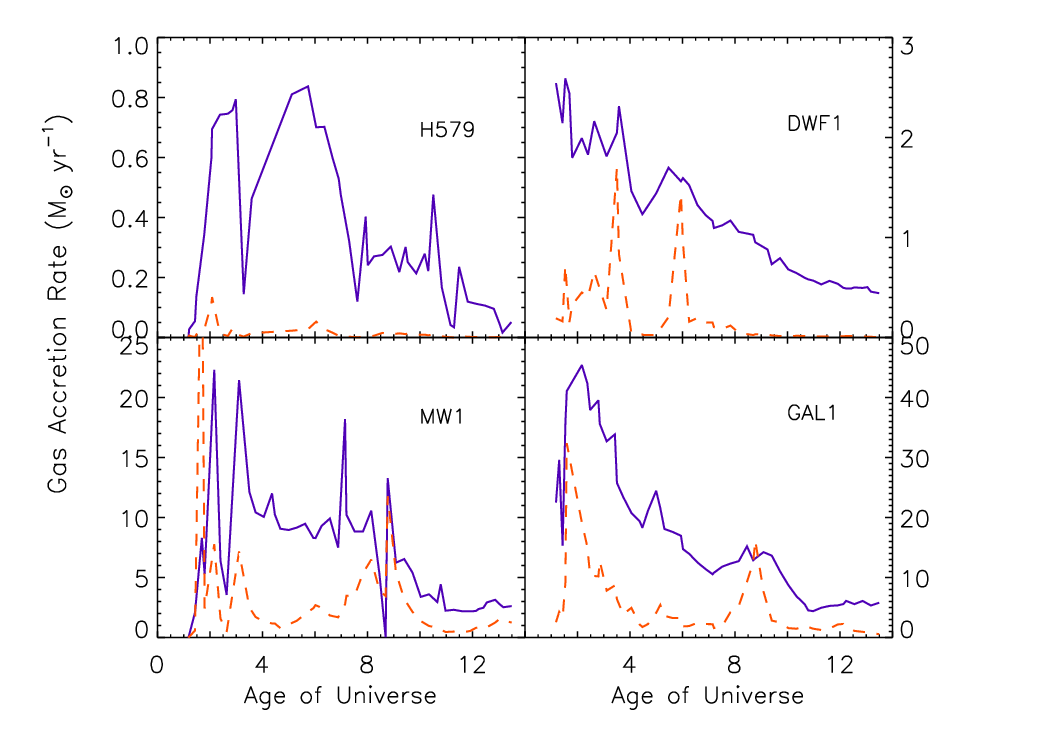}
 \caption{
This plot shown the gas accretion rates at the virial radius
for ``smooth'' and ``clumpy''
accretion as a function of time for each of the four main galaxies.
The solid, purple line shows the mass in gas that has been accreted
smoothly, while the dashed, red line shows the mass in gas that has
been accreted from other galaxy halos.  The present epoch, $z$ = 0,
occurs at 13.7 Gyr.  Narrow, rapid increases in the accretion rate
are characteristic of mergers.}
\label{fig3} 
\end{figure*}

Fig.~\ref{fig3} shows the accretion rate of gas to the virial radius
as a function of time for each of the four main galaxies. 
Lines denoting smoothly accreted gas and ``clumpy'' accretion gas are 
shown in purple (solid) and red (dashed), respectively.  Major 
mergers are conspicuous for all but the lowest mass galaxy, H579, for 
which the last major merger occurred at a redshift
greater than 5.  The obvious result here, again, is that smooth 
accretion of gas that has never belonged to another galaxy halo
dominates the gas accretion history of all of these disk galaxies.
Fig.~\ref{fig3} demonstrates that this is true at all redshifts, and 
not just for the cumulative case that is shown in Fig.~\ref{fig1}.

\subsection{Shocked vs Unshocked Gas}
\label{decomp}

Having now identified two subsets of gas accretion to the virial 
radius, ``smooth'' or ``clumpy,'' we now focus on the gas that has 
been smoothly accreted.  In the classic analytic picture of disk galaxy 
formation, this smoothly accreted gas will be shocked as it enters 
the virial radius of the galaxy.  DB06 showed, instead, that there 
is a critical mass below which galaxies are unable to support a stable 
shock at the virial radius.  If a galaxy is below this mass, then gas 
may not shock until 
it reaches the disk, where the density contrast is high.  In massive 
galaxies, even after a stable shock develops and propagates out through
the galaxy's halo, filaments of cold gas may penetrate inside the shock
radius.  These filaments may shock when they reach a radius closer to
the disk.

\subsubsection{Shock Definitions}

Smoothly accreted gas is followed until it reaches a radius of 30 
kpc\footnote{All distances given in this work are physical, unless 
specified as comoving.} from the galaxy center (or R$_{vir}$ at high 
$z$, whichever is smaller).  We stop following gas particles at 30 
kpc to avoid contamination from feedback effects that occur within our 
galaxy disks that also lead to a strong increase in entropy and 
temperature, mimicking the shocks we wish to identify.  Similarly, we 
exclude ``clumpy'' gas from consideration in the remaining analysis, as 
feedback from merger induced star formation may mimic shocks. 

By definition, a particle undergoes an entropy increase when it encounters
a shock.  Previous work that searched for shocked gas within simulations 
used a temperature criterion rather than an entropy criterion.  Our high 
resolution simulations allow us instead to search for the entropy increase 
associated with shocks.  We searched all smoothly accreted gas for an 
entropy jump that corresponded to a Mach number of 3 or more.  However, 
we found that not all of this shocked gas was associated with an increase 
in temperature to near the virial temperature of the halo (we discuss in 
Section~\ref{entropy} that this is unlikely to be a resolution effect).  
For the purposes of this work, we are most interested in comparing to 
semi-analytic models of galaxy formation, which adopt the assumption 
that gas starts at the virial temperature of the halo before cooling to the 
disk.  Thus, our adopted definitions below include a temperature criterion. 

At each time step, we determine the densities and temperatures 
for each gas particle and compare to the properties of the main halo at 
that step.  In the limit of a strong shock, it follows (see also DB06) 
from the Rankine-Hugoniot shock jump conditions for a singular isothermal 
sphere: 
\begin{equation}
\rho_{shock} = 4\rho_0 
\label{eq:1}
\end{equation}
and
\begin{equation}
T_{shock} \ge 3/8 T_{vir},
\label{eq:2}
\end{equation}
where $\rho_0$ is the gas density prior to encountering the shock, 
$\rho_{shock}$ is the post-shock density, and T$_{vir}$ is the virial 
temperature of the galaxy halo that the gas encounters.  

We require that a gas particle must reach 3/8 T$_{vir}$ to be labeled 
as ``shocked.''  In practice, however, due to the heating of the 
intergalactic medium (IGM) by our UV background, 3/8 T$_{vir}$ of low mass 
halos is below the temperature floor of the IGM.  Alternatively, this 
means that the entropy of these halos is below the entropy floor of 
the IGM (discussed further in the next section), where we adopt the 
entropy definition of S = log$_{10}$(T$^{1.5}$/$\rho$).  Thus, for a 
particle to be labeled as ``shocked,'' it must reach a minimum entropy 
threshold in addition to a temperature (3/8 T$_{vir}$) threshold.  This 
entropy criterion requires
\begin{equation}
S_{shock} \ge log_{10}[(3/8 T_{vir})^{1.5}/4\rho_0].
\label{eq:3}
\end{equation}

For each smoothly accreted gas particle at each time step we ask 
what minimum change in entropy, $\Delta$S, is required for the particle 
to go from its initial entropy, S$_0$ at step {\it t}, to S$_{shock}$ 
by the next time step, {\it t}+$\delta$t (where, again, $\delta$t is 
$\sim$320Myr at most redshifts).  If the entropy at the next time step 
is in fact $\geq$ S$_{shock}$, and the temperature of the particle at 
that next time step is $\geq$ 3/8 $T_{vir}$, we count that particle as 
having undergone a shock.  

In summary, to identify shocked gas we require that a gas particle 
undergoes a strong increase in both entropy and temperature, using two 
simple rules: 
$\Delta$S $\geq$ S$_{shock}$-S$_0$, and T$_{shock}$ $\geq$ 3/8 $T_{vir}$.
Gas particles which meet these criteria are labeled as ``shocked,'' and 
the remaining smoothly accreted gas particles are then identified 
as ``unshocked.''  If filaments undergo either strong shocks or a series 
of weak shocks within the halo, our definitions will identify both of 
these processes. 

Finally, we restrict our label of ``shock'' to those particles that shock 
at the time they enter the main galaxy's virial radius, or afterward.  We 
note that many of our particles do in fact undergo 
entropy jumps which meet our definition of shocks at earlier times.  
It is common for these earlier shocks to occur while the gas particle 
is still at 
distances of several R$_{vir}$ from the main galaxy, and at times many 
Gyr before they are ultimately accreted.  As we presume these shocks have 
nothing to do with galaxy accretion (our focus here), we ignore them (and 
note that these shocks are likely to occur when the UV background turns on, 
or in collapse in filaments).  

Fig.~\ref{fig1} shows the results of our search.  The smoothly accreted 
gas has been divided into ``shocked'' (red) and ``unshocked'' (blue),  
based on the temperature and entropy criteria described above.  
Immediately obvious is that there is a strong trend for the amount of
shocked gas within a galaxy to increase with virial mass. 
The last two columns of Table~\ref{accr} quantify the fraction of shocked 
and unshocked gas mass to the total {\it smoothly accreted} gas mass. 
When considering only the fraction relative to smoothly accreted gas 
(rather than all gas), the trend with mass in the table is even stronger 
than originally appears in  Fig.~\ref{fig1}.

\subsubsection{A Minimum Halo Mass for Shocking}

\citet{BD03}, K05, BD06, K08, and OPT08 found that there is a minimum halo 
mass required to support a stable shock.  
Similarly, a minimum halo mass to shock is built into our definition,  
which can be deduced from Eqn.~\ref{eq:3}, as follows.

Gas particles in the intergalactic medium will be at the 
mean baryonic density for that redshift, $\rho_{b}$.  However, their 
temperature will be set by the ionizing UV background 
radiation.\footnote{The temperature of our UV background was quantified 
by running a simple gas and dark matter run with no star formation, 
no gravity, and cooling on.}  Thus, these particles have some entropy, 
S$_{mean}$.  If particles with S$_{mean}$ are to shock and reach 
S$_{shock}$ as they enter the galaxy's virial radius, their densities 
must jump to 4$\rho_{b}$, and their temperature must be 3/8 T$_{vir}$.  
For low halo masses, though, with low T$_{vir}$, S$_{shock}$ is less 
than S$_{mean}$.  If this is the case, our shock definitions would lead 
to the false conclusion in which every accreted gas particle appears to 
be shocked.  Effectively, the sound speed in the infalling gas is 
comparable to the infall velocity so there is no opportunity for a 
supersonic-subsonic transition.  Clearly, we must require S$_{shock}$ 
$>$ S$_{mean}$ for gas to be physically shocked.  

\begin{figure*}
\plotone{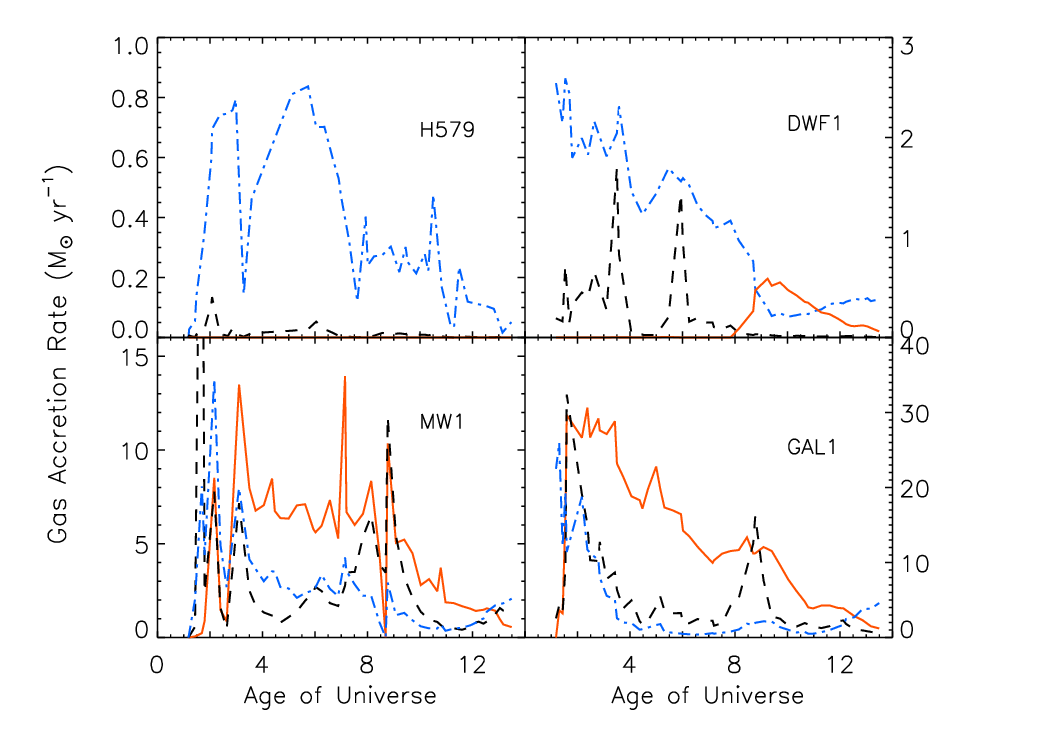}
 \caption{This plot is the same as Fig.~\ref{fig3} except that the
``smooth'' gas accretion rate has been divided further into ``shocked''
(red, solid line) and ``unshocked'' (blue, dot-dashed line) accretion.
The ``clumpy'' gas accretion rate is shown by the black, dashed line.
The present epoch, $z$ $=$ 0, occurs at 13.7 Gyr.  }
\label{rate2} 
\end{figure*}

This requirement then leads to a minimum halo mass that is capable of 
shocking, as a function of redshift.  For our lowest mass galaxy, H579, 
S$_{shock}$ never becomes greater than S$_{mean}$.  Hence, all of the 
smoothly accreted gas in H579 is identified as unshocked gas.  
DWF1 does not shock gas until it reaches a halo mass of 
1.1$\times$10$^{11}$ \Msun ~at $z$ = 0.6.  This late shock development 
is seen in detail in Fig.~\ref{rate2}, which is the same as 
Fig.~\ref{fig3}, but now with smoothly accreted gas divided into shocked 
and unshocked gas.  Meanwhile, our two most massive galaxies (MW1 and 
GAL1) reach large enough masses to shock very early on.  For MW1, this 
happens at $z$ = 4 when the galaxy is 8.9$\times$10$^{10}$ \Msun.  GAL1 
is 1.6$\times$10$^{11}$ \Msun ~at $z$ = 6 (as far back as we trace gas), 
at which time it is already capable of supporting shocks.

Although these four galaxies do not provide a statistical sample, the 
mass at which their shocks exist are in good agreement with previous 
results.  For DWF1 and MW1, the masses listed in the previous paragraph 
are the masses at which a shock is capable of developing, and thus 
when a hot halo might develop. GAL1 is already capable of shocking 
at $z$ = 6, so this mass is an upper limit to when a stable shock develops.  
OPT08 show that shocking near the disk radius begins to occur when 
dark matter halos reach masses of a few$\times$10$^{10}$ \Msun ~to 
10$^{11}$ \Msun (see their figure 4), in good agreement with the masses 
found for our galaxies.  We note that the shock definition of these 
authors differs from ours, but find below (Section~\ref{temp}) that our 
two definitions yield similar results.  

These four galaxies were chosen to have their last major mergers at 
$z >$ 1.5.  If we had instead examined galaxies with late major mergers, 
they may not cross the shock threshold until low redshifts, and have 
even higher fractions of unshocked gas accretion.  Our galaxies here 
assemble the majority of their mass early, and then quietly accrete 
gas afterward, meaning that they effectively cross the mass threshold 
for stable shocks early.  Thus, by adopting galaxies with similar merger
histories, we guarantee that we are comparing galaxies across a range
in mass, but with other properties being similar.

Because of their range in final mass, each galaxy crosses the shock
threshold at differing times.  Galaxies that cross this threshold sooner,
i.e., more massive galaxies, therefore have a longer period of time to
accrete shocked gas.  This extended time capable of shocking
is partly responsible for the well defined trend of accreted shock fraction
with increasing mass, in that these galaxies have similar histories and
thus display this trend sharply.  However, larger volume cosmological
simulations that are capable of statistically sampling a large number of
galaxies also show this trend (K05, K08, OPT08).

\subsubsection{The Temperature History of Shocked and Unshocked Gas}

\begin{figure*}
\plotone{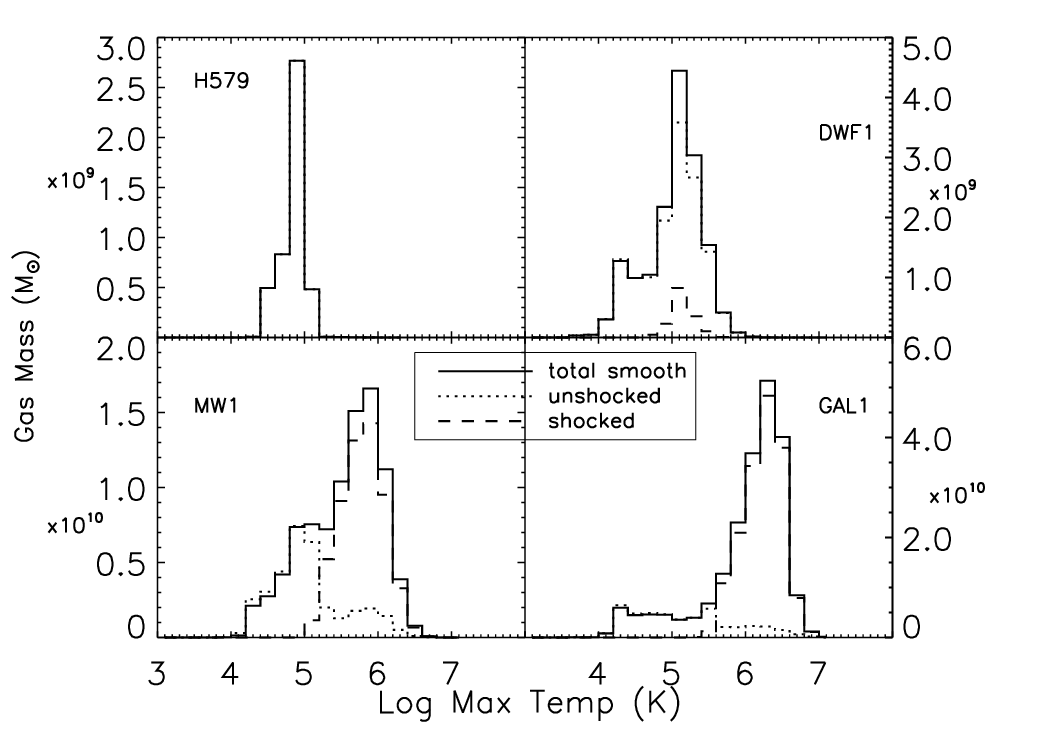}
\caption{
This figure shows histograms of the maximum temperature reached for 
all of the {\it smoothly} accreted gas for each of our four simulations 
(solid line).  The dotted lines show the component of smoothly accreted 
gas that remains unshocked after being accreted to the galaxy, while 
the dashed line shows the gas that is shocked after entering the galaxy.  
} 
\label{fig7}
\end{figure*}

Fig~\ref{fig7} shows histograms of the maximum temperature attained by 
all the gas that is smoothly accreted by our four galaxies.  The 
temperature shown here is the maximum temperature the particle reaches 
before we stop tracing it (i.e., until the particle reaches 30 kpc of 
the galaxy's center).  The solid line shows all of the smoothly 
accreted gas, while the dotted lines show the component of that 
smooth accretion that has remain unshocked, and the dashed lines 
show the shocked component. 

In general, the shocked/unshocked gas is well divided into hot/cold as 
well.  H579 has no shocked gas, but for our two most massive galaxies, 
MW1 and GAL1, the two components are well separated in temperature. 
This separation is not altogether surprising, as part of our requirement 
for being identified as shocked is that the particle reach a minimum 
temperature (3/8 T$_{vir}$) as well as have a sudden entropy increase.  
The exception is DWF1.  While the unshocked gas extends to much lower 
temperatures than the hot gas, it overlaps the shocked gas.  This result 
is due to the requirement mentioned above that the shocked entropy of the 
galaxy must be above the ambient, intergalactic entropy (S$_{mean}$, set 
by the mean baryon density and the UV temperature floor).  DWF1 does 
not cross this threshold until $z \sim$0.6.  Consequently, there exists
gas that reaches temperatures above 3/8 T$_{vir}$ of DWF1 at higher 
redshifts, and is not included as shocked.

It is difficult to say immediately whether this gas {\it should} be 
counted as shocked gas.  Note that even the two most massive galaxies, 
which are capable of supporting a shock over basically their entire 
traced histories, include a portion of gas that reached high temperatures 
(above a few 10$^5$ K) yet is labeled as unshocked because it does not 
undergo a strong entropy increase.  Previous works on this subject have 
attempted to identify shocked and unshocked gas by using only a 
temperature cut (e.g., K05 and OPT08 used a cut of 
2.5$\times$10$^5$ K).  In that case, our hot, unshocked gas would have 
been included as shocked gas.  We will investigate this alternative 
definition to search for hot and cold flows later in the discussion 
(Section~\ref{temp}).   

\begin{figure}
\centering
\plotone{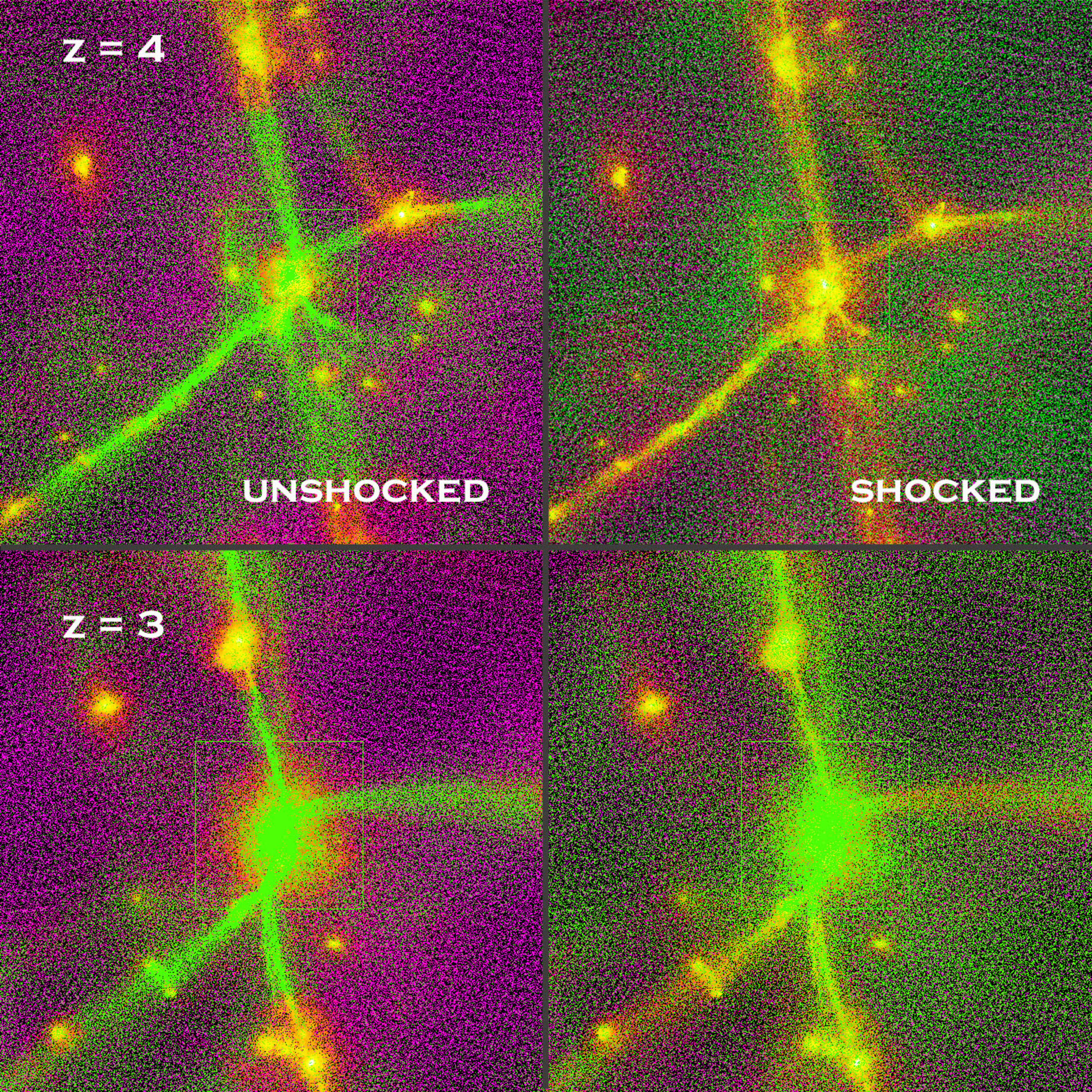}
\caption{Images from our MW1 simulation at redshifts 4 and 3, identifying
gas that will be smoothly accreted to the galaxy as either unshocked or
shocked gas.  {\it Left column:} The distribution of particles that will 
be (or have been) accreted but remain unshocked are marked in green.  
{\it Right column:} Particles that will be (or have been) shocked as they 
are accreted are shown in green.  The underlying colors represent a gas 
density map, for reference,
with black being least dense and white being the densest structures.
The frames are centered on the main MW1 progenitor at each of the
redshifts shown.  Each frame is $\sim$1 comoving Mpc on a side.  Faint, 
green boxes indicate R$_{vir}$ at each time.  }
\label{highz}
\end{figure}

Despite the small inclusion of hot gas in the unshocked definition, we 
find that our definitions do an excellent job of separating gas spatially, 
into gas accreted along preferred directions in filaments 
(unshocked) and gas accreted from a more spherical distribution (shocked).  
This distinction was also found by K05 and OPT08, using their temperature 
definition.  Fig.~\ref{highz} demonstrates this for our MW1 galaxy.  At 
$z$ = 4 and 3, the 
left hand column shows in green the gas particles that will eventually 
be (or have already been) accreted to the galaxy but remain unshocked, 
while the right hand column shows in green those gas particles that will 
be accreted (or already have been) to the galaxy and shocked.  The 
underlying colors are simply a gas density map, for reference (black is 
least dense, white is densest structure).   Each frame is $\sim$1100 
comoving kpc on a side.  Shown as a faint green box is the virial radius 
of our MW1 galaxy at each time.  At $z$ = 4, the galaxy has only just 
crossed the entropy threshold to allow shocked gas, so that it is only
accreting unshocked gas at this step.  However, by $z$ = 3 the galaxy is 
capable of accreting shocked gas, but is still accreting a substantial 
amount of unshocked gas in filaments penetrating deep inside of the virial
radius.  This demonstrates that the two modes of accretion can 
occur simultaneously at high $z$.  

The immediately obvious feature in Fig.~\ref{highz} is the clear 
distinction in spatial distribution at these high redshifts, before 
most of the gas has been accreted.  The unshocked gas is collapsing into 
filaments that are feeding the galaxy's growth.  Because these filaments 
are capable of penetrating deep inside the virial radius, this cold gas 
is being delivered much closer to the galaxy disk (see also section 
\ref{adiabatic}, and hence appears to occupy a smaller spatial scale 
than the shocked gas.  The gas that will later be accreted and shocked, 
on the other hand, has a much more uniform or spherical distribution.  The 
shocked gas develops into a hot halo, much further from the disk.

\section{The Growth of the Stellar Disk}
\label{diskgrowth}

Now that we have identified how gas gets to the galaxy, and whether it 
shocks or remains unshocked as it approaches the disk, we wish to take 
it one step further and discover what role the various gas accretion 
mechanisms play in building the galaxy stellar disk.  

To identify disk particles, the galaxies are first aligned so that the 
disk angular momentum vector lies along the {\it z}-axis.  J$_z$/J$_{circ}$ 
is calculated for each star particle in the galaxy, where J$_z$ is 
the angular momentum in the {\it x-y} plane, and J$_{circ}$ is the momentum 
that a particle would have in a circular orbit with the same orbital energy 
\citep[see also][]{Abadi03}.  Disk stars are identified as those having near 
circular orbits, so that J$_z$/J$_{circ}$ $>$ 0.8. 
For a full discussion on our disk identification, see Brook \etal (2008, 
in prep).

\begin{figure*}
\plotone{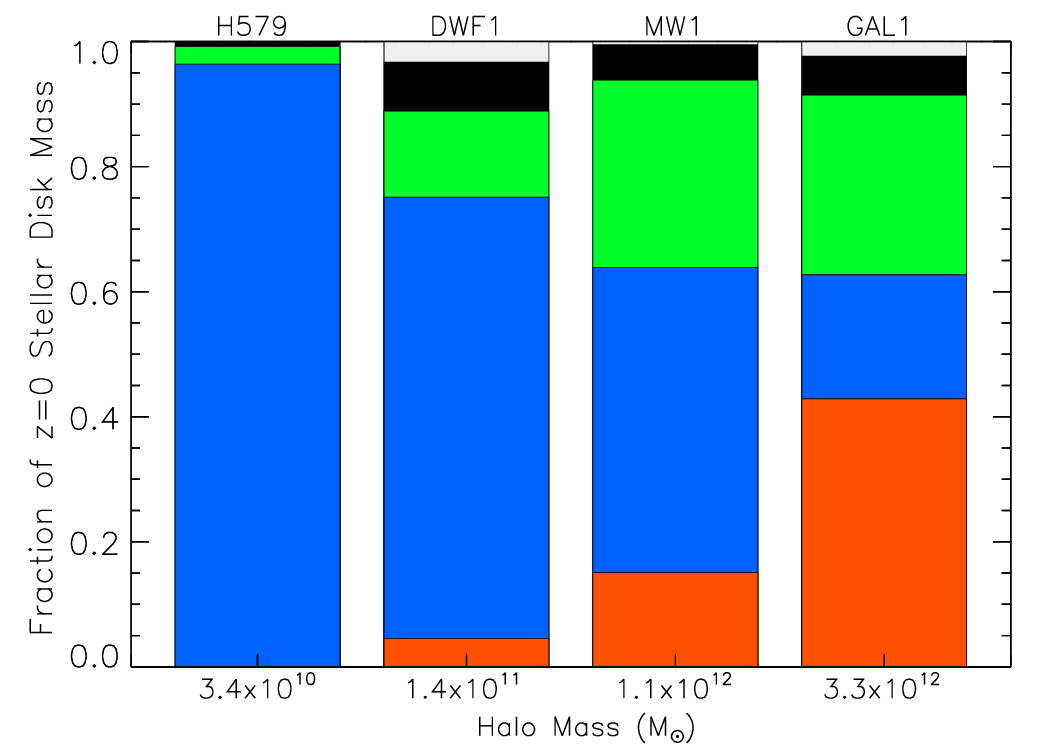}
\caption{Cumulative fraction of $z$ = 0 stellar disk mass by source, for 
each of our four galaxies.  {\it red}: fraction of stellar disk mass that
formed from shocked gas; {\it blue}: fraction of stellar disk mass that
formed from unshocked, smoothly accreted gas particles that never belonged 
to a satellite galaxy before joining the main galaxy; {\it green}: fraction 
of the stellar disk mass that formed from ``clumpy'' accretion, i.e., gas
that at some point was part of another galaxy halo; {\it black}: fraction
of stellar disk mass that was accreted directly as stars from other halos;
{\it white}: the fraction of stellar disk mass that formed from gas already
in the main galaxy at $z$ = 6.  The total halo mass (in \Msun) of each 
galaxy is listed below its respective bar.  }
\label{fig8}
\end{figure*}

Fig.~\ref{fig8} shows the 
fraction of stellar disk mass at redshift zero that has formed from gas of
various sources, for each of our four main galaxies.  Again, the total
halo mass of the four galaxies is listed below their respective bar, for
reference.  The total of the blue and red bars shows the fraction of stellar
mass in the disk that formed from smoothly accreted gas, with red indicating
the fraction that shocked as it entered the galaxy, and blue indicating
the fraction that remained unshocked.  The green bar shows the fraction
of the stellar disk mass that formed from ``clumpy'' accretion, i.e., gas 
that at some point belonged to another galaxy halo.  Finally, the black bar 
shows the fraction of stellar disk mass that was accreted directly {\it as 
stars} from other galaxies, while the white shows the fraction of stellar 
disk mass that formed from gas that was already in the galaxy at $z$ = 6. 

Fig.~\ref{fig1} showed that smoothly accreted gas dominates over clumpy
accretion for all four of our galaxies.  Fig.~\ref{fig8} shows 
that this smooth gas goes on to form stars, so that the majority of the disk
is composed of stars formed from smoothly accreted gas rather than from gas
accreted in mergers or {\it stars} accreted in mergers.  Table~\ref{disk}
is similar to Table~\ref{accr} but for the stellar disk at redshift zero.
Unlike Table~\ref{accr}, in which the fraction of smooth and clumpy gas totaled
to 1.0, in Table~\ref{disk} the fractions of stars formed from smooth,
clumpy, and ``early'' gas now sum to 1.0, where ``early'' gas is that
which was already in the galaxy at $z$ = 6, i.e., as far back as we trace
the history of our gas particles.  However, the fractions of stars that
formed from shocked or unshocked gas (the last two columns in Fig. 
\ref{fig8}) again sum to 1.0 (representing the total of stars formed 
from smoothly accreted gas).

It can now be seen by comparing Tables~\ref{disk} and \ref{accr} that
the amount of shocked gas that makes it to the disk to form stars is
lower than the total amount of shocked gas that is accreted to the galaxy.
In fact, Table~\ref{disk} shows that for all but our most massive galaxy,
the stellar disk mass is dominated by stars formed from unshocked, smoothly
accreted gas.  This domination of unshocked gas in building the stellar
disk is even true for MW1, despite the fact that the
majority of smoothly accreted gas to MW1 gets shocked.

\begin{deluxetable}{lcccc}
\tablecaption{Disk Star Forming Properties \label{disk}}
\tablewidth{0pt}
\tablehead{
\colhead{galaxy} & \colhead{f$_{\mathrm{smooth}}$} & \colhead {f$_{\mathrm{clumpy}}$} &
\colhead{f$_{\mathrm{shock}}$} &  \colhead{f$_{\mathrm{unshock}}$} }
\tabletypesize{\footnotesize}
\startdata
H579    & 0.96 & 0.03 & 0.00 & 1.00 \\ 
DWF1    & 0.75 & 0.14 & 0.06 & 0.94 \\ 
MW1     & 0.64 & 0.30 & 0.24 & 0.76 \\ 
GAL1    & 0.63 & 0.29 & 0.69 & 0.31 \\ 
H277    & 0.67 & 0.23 & 0.32 & 0.68 \\ 
MW1.lo  & 0.70 & 0.28 & 0.28 & 0.72 \\
\enddata
\tablecomments{The first two columns, f$_{\mathrm{smooth}}$ and 
f$_{\mathrm{clumpy}}$, are the fractional quantities with respect to 
the total mass in stars in the galaxy disk at $z$ = 0.  Here, 
f$_{\mathrm{clumpy}}$ includes only stars that formed from gas
accreted from other halos (and not stars accreted directly as stars from
satellites).  By definition, f$_{\mathrm{smooth}}$, f$_{\mathrm{clumpy}}$, 
and f$_{\mathrm{early}}$ (not listed here) sum to 1.0.   The last two 
columns, f$_{\mathrm{shock}}$ and f$_{\mathrm{unshock}}$, are the fractional 
quantities of stars that formed from shocked gas or unshocked gas with 
respect to the total stellar disk mass that formed from {\it smoothly} 
accreted gas that is in the galaxy at $z$ = 0 (and sum to 1.0).  }
\end{deluxetable}

\begin{figure*}
\plotone{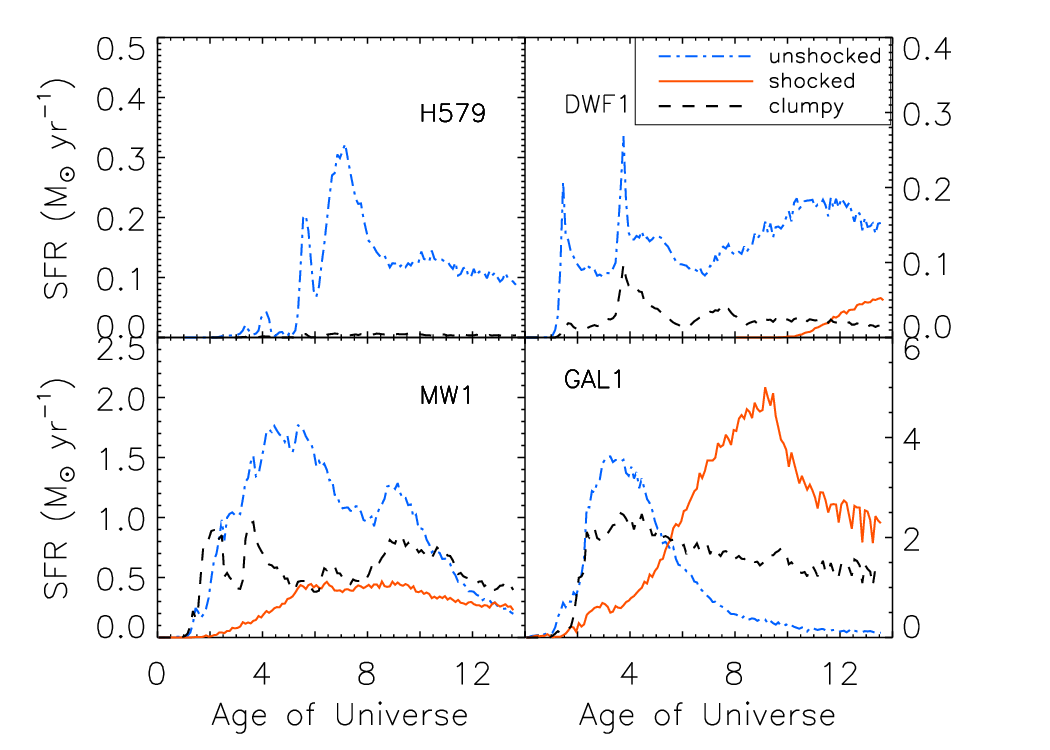}
\caption{The star formation history for those stars that are in the disk 
of each galaxy at $z$ = 0.  The star formation rates are broken down 
according to the accretion mode of the gas progenitor particles.  Stars 
that formed from ``clumpy'' accreted gas are represented by the black, 
dashed line.  The star formation rate of stars formed from smoothly 
accreted, unshocked gas is shown by the blue, dot-dashed line.  Finally, 
the star formation rate for stars formed from smoothly accreted, shocked 
gas is shown in red (solid line).  $z$ = 0 occurs at an age of 13.7 Gyr.  }
\label{diskgrow}
\end{figure*}

\subsection{The Early Growth of the Stellar Disk}
\label{early}

The importance of unshocked gas in the growth of the stellar disk is 
shown more explicitly in Fig.~\ref{diskgrow}, which shows the star
formation histories of those stars that are in the galaxy disk at
redshift zero, for each of the four main galaxies (note, not the SFH
of the entire galaxy).
Here we can see that the disk SFR is, in fact, dominated at basically 
{\it all times} by stars forming from smoothly accreted, unshocked gas, 
for all but the most massive galaxy, GAL1.  Even in GAL1, 
unshocked gas dominated the growth of the stellar disk at an 
early period (to $z \sim$1); only later does shocked gas begin to 
dominate.

It is interesting to compare Fig.~\ref{diskgrow} to Fig.~\ref{rate2}, 
which shows the gas accretion rate onto the virial radius with time.
Note that in no case is the gas accretion rate exactly balanced by the 
star formation rate, though again we emphasize that we are only including 
disk SF here (and excluding, for example, the SFR of the bulge).
Our two most massive galaxies are particularly interesting.
Fig.~\ref{rate2} shows that their gas accretion histories are generally
dominated by smoothly accreted, {\it shocked} gas.  However, the star
formation history of the MW1 disk is instead dominated by smoothly
accreted, {\it unshocked} gas.  Star formation from gas that has been
shocked initially increases steadily with time, until about 5 Gyr, after
which the disk SFR due to this shocked component still remains relatively
low, and eventually declines.  GAL1 is a dramatic example of this.  Shocked
gas dominates the gas accretion history of the galaxy at essentially all
times in the traced history.  Yet again the initial star formation of the
disk is instead dominated by unshocked gas accretion, while the star
formation from shocked gas slowly increases with time, before peaking at
about 9 Gyr and again declining.

This delay in star formation for the shocked gas is not surprising.  As 
this gas reaches higher temperatures (and entropies), it has much
longer cooling times, slowing the rate at which it can cool to the disk
and form stars.  As the galaxy grows in time and reaches even higher
virial temperatures, the radius at which shocking occurs increases 
(discussed further in Section~\ref{adiabatic}).  This increase in shock 
radius will potentially make later accretion to the disk even more 
difficult, due to even longer cooling times.

We have quantified the different delay times (the time between accretion 
to the virial radius and star formation in the disk) of the two types 
of smooth gas accretion.  We find those stars that form from shocked 
and unshocked gas, and measure the average time between a gas particle's 
accretion to the virial radius and the time when it begins to form stars.
This delay is shorter for the unshocked gas.  The typical
delay time for unshocked gas in our MW1 and GAL1 galaxies is 1-2 Gyr.
For shocked gas, the delay times in MW1 are typically 3-4 Gyr, while for
the hotter GAL1 the times are longer, 4-6 Gyr.

Note that these delays are NOT intended to be representative of the cooling 
times of the particles.  The delay between accretion and star formation 
includes both the cooling time of the gas particle onto the disk, the 
natural time the gas might spend in the disk before producing stars, and 
subsequent delays in SF due to feedback processes in the disk, which vary 
as a function of galaxy mass \citep{Brooks07}.  
In the two lowest mass galaxies, the delay time for unshocked 
gas is typically longer than for our two highest mass galaxies (3-4 Gyr
rather than 1-2 Gyr) due to the increasing importance of feedback 
effects at lower masses.
 
The main conclusion, however, is that we expect longer cooling times for
shocked gas compared to clumpy or unshocked gas.  Thus, clumpy and unshocked
gas will be able to make it to the disk to form stars at a faster rate than
the shocked gas.  In the next two sections we discuss the impact of cold 
flow (unshocked) gas accretion on disk formation.

\subsection{Observable Properties}
\label{observables}

\begin{figure}
\plotone{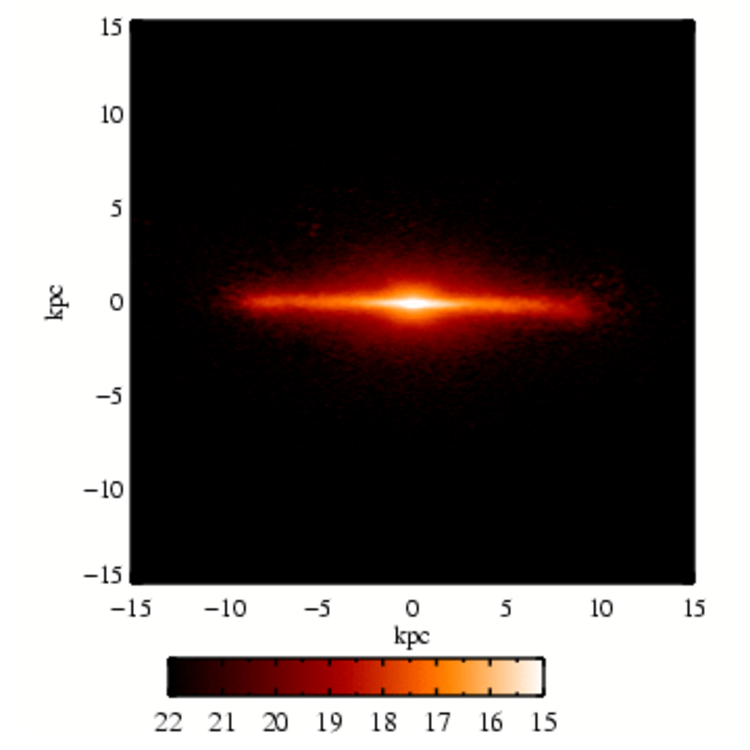}
\caption{
The rest frame $B$ band surface brightness images of GAL1 at $z$ = 1, from
{\sc Sunrise} (in the case with dust scattering turned off).  Note the obvious 
disk, even at this redshift.}
\label{gal1}
\end{figure}

Fig.~\ref{gal1} shows the rest frame $B$ band surface brightness for 
GAL1 at $z$ = 1.  This image was created using {\sc Sunrise}, a Monte 
Carlo radiative transfer code \citep{sunrise}.  The growth of the disk in 
GAL1 is dominated by unshocked gas accretion prior to $z$ = 1, which 
clearly results in an observable disk by this redshift.  
In a future paper we will investigate the evolution of disk
sizes since $z$ = 1. 

\begin{figure*}
\plotone{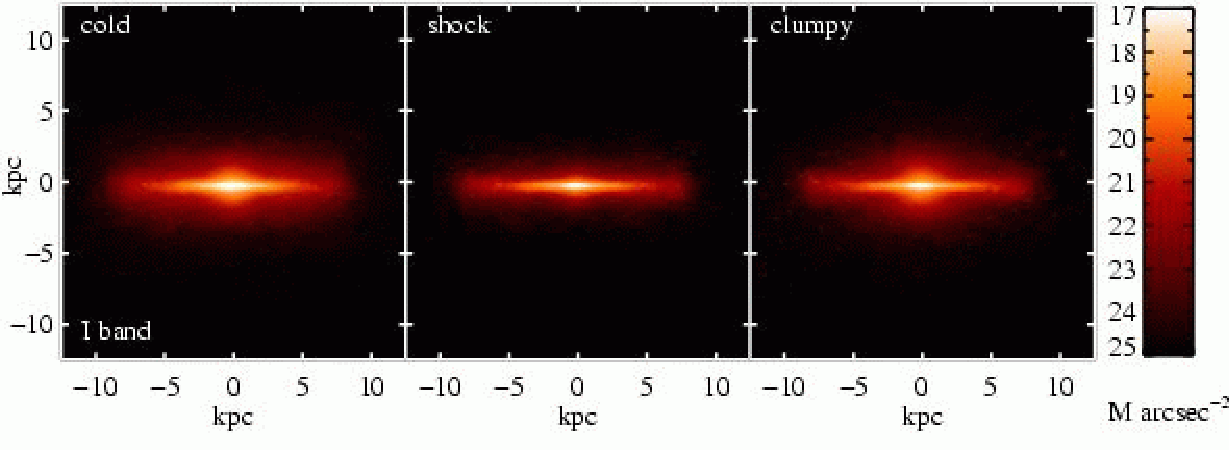}
\caption{
$I$ band surface brightness images from {\sc Sunrise}, for our MW1 galaxy
at $z$ = 0 (again with dust scattering turned off).
{\it Left:} stars formed from unshocked, smoothly accreted gas.
{\it Middle:} stars formed from shocked, smoothly accreted gas.
{\it Right:} stars formed from clumpy accretion gas.
} 
\label{fig5}
\end{figure*}

Fig.~\ref{fig5} shows {\sc Sunrise} images of the $I$ band surface
brightness of stars at $z$ = 0 that form from the various accretion modes
in our MW1 simulation.  The left panel shows stars formed from unshocked, 
smoothly accreted gas; the middle panel shows stars that formed from 
shocked, smoothly accreted gas; the right panel shows stars that formed 
{\it in the galaxy} from gas that had been accreted from other halos 
(i.e., clumpy gas accretion).  These panels show that the stars formed 
from the clumpy and unshocked, cold gas are found in both the stellar disk 
and bulge component, while the stars that are able to form from shocked gas 
are instead predominantly in a disk.  This disk formed from shocked gas is
also thinner than the disk created by the unshocked or clumpy gas.  The 
scale height of the shocked disk is $\sim$25\% smaller than the scale 
height of the unshocked disk.

The morphology indicated by Fig.~\ref{fig5} can be explained in terms
of the accretion times of the various gas components, and the subsequent
cooling rate of that gas.  The earliest disk stars are predominantly
formed by unshocked and ``clumpy'' gas, which dominate the accretion rate
to the galaxy at the earliest times.  This gas can rapidly cool and
form stars, but at these early times the galaxy is still undergoing a series
of mergers (major and minor) that will both heat the forming disk (leading
to a larger scale height for the stars created from unshocked and
clumpy gas) and lead to bulge formation.  The shocked gas, on the other
hand, is both the dominant mode of accretion at later times and takes 
substantially longer to cool to the disk after accretion.  The shocked 
gas will therefore settle into the disk after the chaotic merger period.  
This late settling allows the shocked gas to cool to a thin disk before 
forming stars \citep [see also][]{brook04}.  Hence, the scale height of the 
stellar component formed from this shocked gas is smaller than the unshocked 
case.

\begin{figure}
\plotone{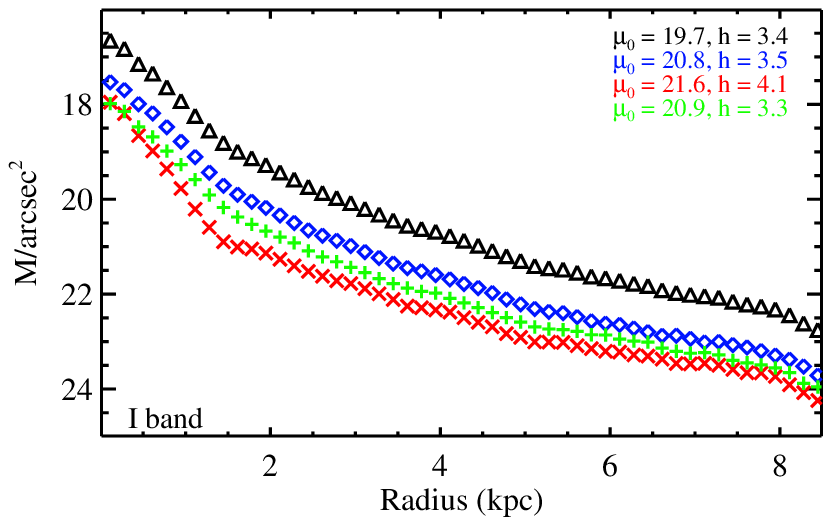}
\caption{
Average surface brightness versus radius for the {\sc Sunrise} images in 
Fig.~\ref{fig5}. Blue 
diamonds show the contribution from smoothly accreted, unshocked gas; red 
X's from smoothly accreted, shocked gas; green +'s from clumpy accretion 
gas; black triangles show the total.  It can be seen that unshocked gas 
dominates the disk magnitudes at all radii, but that the shocked gas 
has a longer scale length (listed as $h$ in the upper right corner).  See 
the text for further discussion.
} 
\label{sb}
\end{figure}

Fig.~\ref{sb} dissects these {\sc Sunrise} images further, showing the 
radially averaged surface brightness profiles for each of these three 
components of gas accretion.  As expected from Fig.~\ref{diskgrow}, 
the stars that form from unshocked gas dominate the surface brightness 
(shown as blue diamonds; clumpy as green +'s; shocked as red X's; total 
as black triangles).  These profiles are fit well by a Sersic bulge 
component plus exponential disk.  

The scale lengths, $h$, for each type of accretion are listed in the
upper right corner of Fig.~\ref{sb}.  The stars formed from shocked gas
have a longer disk scale length (4.1 kpc) than those formed from unshocked
(3.5 kpc) or clumpy gas (3.3 kpc).  Although the picture is confused
slightly due to the longer cooling times of the shocked gas, this scale
length difference may not be surprising given the differing accretion times
for these two components.  Although there is still a low level of unshocked
gas accretion at the virial radius all the way to redshift zero, the
majority of the unshocked gas is accreted prior to $z$ = 2 when the mean
angular momentum is expected to be lower, and when chaotic merging will
heat the disk.  After this time, shocked gas dominates the gas accretion
to the halo, when the mean angular momentum is expected to be higher, and
will therefore form a disk with a longer scale length.  The angular
momentum content of this gas will be investigated in detail in a future
paper.

\subsection{Comparison with SAMs}
\label{SAMs}

An important point to draw from the previous two sections is that, 
particularly as seen in our two most massive galaxies, a significant 
amount of star formation is possible in disks prior to $z$ = 1 (when 
the universe was 6 Gyr old).  As we have shown above, this early star 
formation is predominantly due to the contribution from unshocked gas 
accretion.  Of course, the disks continue to grow since $z$ = 1, but an 
existing disk is in place already by that time.  This is in accord with 
observations that find that large galaxy disks must be assembled prior 
to $z$ = 1 \citep{vogt96, roche98,lilly98,simard99, labbe03, rav04, 
ferguson04,trujillo04,barden05, sargent07, melbourne07,kanwar08}, or 
even by $z$ = 2 \citep{sins1,sins2,genzel08,D08}.

Theoretical models of disk formation suggest instead that large disks 
must have formed since $z$ = 1 and hence undergone a large change in 
scale length since then, \citep{mmw,mao98,vdb98}.  Even the most 
up-to-date models still suggest a discrepancy with observational 
results \citep{somerville08}, with observed disks being larger than 
predicted at $z$ = 1.  We will address disk sizes in a future paper,
as it also requires a detailed examination of the angular momentum content 
of the accreted gas.  We focus here on the age of the stellar disk. 
 
It is particularly interesting to examine the stellar disk growth for 
our two most massive galaxies, which have been dominated by shocked gas 
accretion over most of the history we trace here.  The unshocked gas 
accreted to these galaxies never reaches a temperature above 3/8 T$_{vir}$. 
SAMs adopt a model in which this gas would first have been heated to 
the virial temperature of the halo as it entered the virial radius, thus 
delaying its ability to cool quickly to the disk.  As demonstrated in 
Fig.~\ref{fig7}, the temperature difference between the shocked and 
unshocked gas can be a factor of 10.  This temperature discrepancy may 
lead to significant differences between the SFRs found here and in SAMs.

In order to quantify the difference in SFRs between SAMs and our 
simulations, we estimate the effect on the disk SFH if the cold, unshocked
gas had instead first been heated to T$_{vir}$.  The resulting change in
the disk star formation rate is not straightforward to assess.  Even our 
cold flow gas does not immediately form stars when it reaches the disk.  
Rather, there is a delay in star formation due to feedback effects in 
the disk.  As found by \citet{Brooks07}, this delay is a function of 
galaxy mass, being longer in low mass galaxies, so that low mass galaxies 
form stars inefficiently compared to higher mass galaxies.

As discussed in Section~\ref{early}, the time between accretion 
to the virial radius and formation of disk stars is much longer for
shocked gas than for unshocked gas.  We find the shocked and unshocked 
gas particles accreted between each output time step interval, and the 
mean difference in formation time for stars that subsequently form from 
that gas.  To quantify the affect of heating our unshocked 
gas to high temperatures, we adopt this average difference in formation 
time between shocked and unshocked star particles, and add this delay to 
the formation times of the star particles born from unshocked gas.  
Although not precise, this method captures the relevant physics, including 
cooling times and suppression of star formation by feedback, and will 
provide a meaningful estimate of the effects of assuming that all disk 
gas has been shock heated.

\begin{figure*}
\plotone{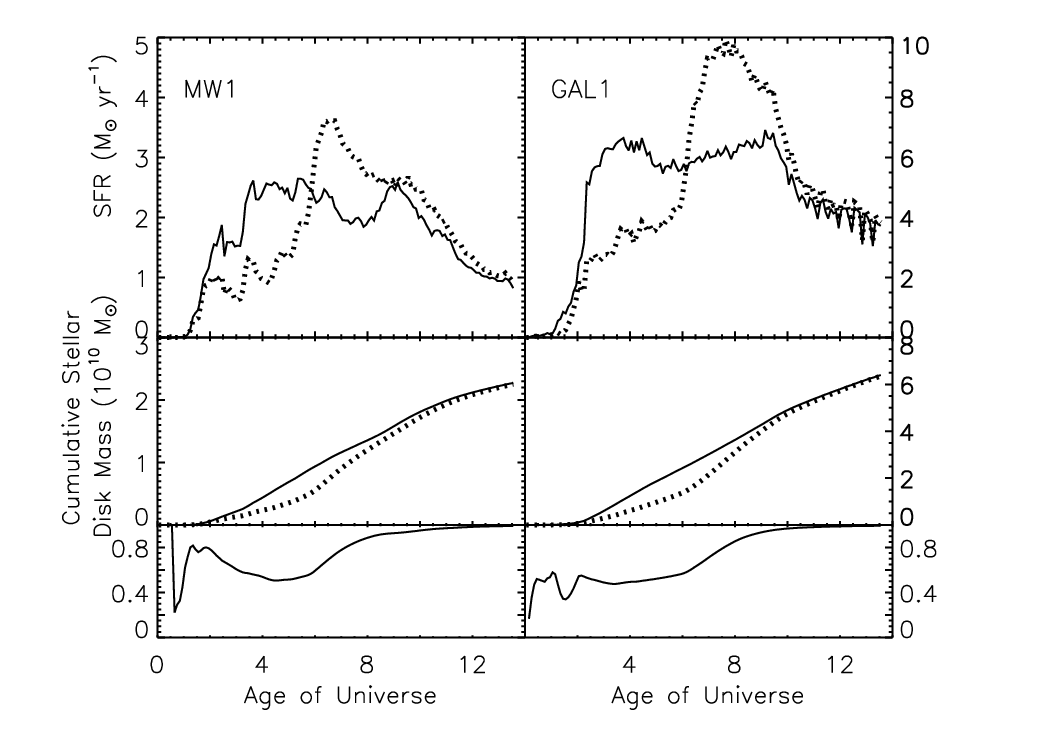}
\caption{
The top panels show the disk SFR in our simulations, as in Fig.~\ref{diskgrow}, 
but now summed together as the solid line.  The dashed line shows the effect 
on the disk SFR if unshocked gas took as long to reach the disk and form 
stars as the shocked gas.  MW1 is shown on the left, and GAL1 on the right.  
The middle panels show the cumulative stellar disk mass in each case.  The 
bottom panels show the difference of the two disk masses shown in the middle 
panel, i.e., the delayed disk mass as a fraction of the disk mass 
from the simulation. }
\label{cooltheory}
\end{figure*}

The result of adding this delay is shown in Fig.~\ref{cooltheory}, for
MW1 and GAL1.  The top panels show the actual disk SFRs from our 
simulations as a solid line, exactly as in Fig.~\ref{diskgrow}, but now 
showing the total disk SFR from unshocked, shocked, and clumpy accreted 
gas together.  The dashed line shows what the resulting SFR would be if 
the unshocked gas had been heated to the same temperatures as the shocked 
gas.  The middle shows the resulting cumulative mass of the stellar disk 
as a function of time, again as actually seen in the simulations (solid 
line) and with the false delay added to the unshocked gas (dashed line).  
The bottom panels show the difference between the two disk masses in the 
middle panel (the delayed disk mass as a fraction of the actual disk 
mass from the simulation).  

Fig.~\ref{cooltheory} demonstrates that if the cold gas had reached 
virial temperatures that the SFR in the disk would have been significantly 
delayed.  Rather than being able to build the stellar disk within the first
few Gyr, the disk does not begin to grow substantially until $z \sim$1.  
The bottom panel shows that the disk is 50\% less massive until $z$ = 1
in the case of strong heating than it actually is in the simulation, where 
the unshocked gas does not approach the virial temperature.

\section{Numerical Issues}
\label{issues}

We discuss in this section the impact of key numerical effects on our 
results. 

\subsection{Time Resolution}
\label{timeres}

The effect likely to have the largest impact on our results is the interval 
between our outputs.  Given our finite step interval of $\sim$320 Myr, it 
is possible that a particle with a short cooling time may radiate 
away its energy quickly enough that we miss an increase in temperature and 
entropy
associated with a shock event (making our shocked fractions a lower limit).  
This omission is more likely to occur for those particles shocking near 
the disk (at $\sim$ 30 kpc) where the densities are much higher (and hence 
cooling times shorter) than at the virial radius.

To investigate the effect of our step size on our results, we have run a 
new simulation with output steps at 1/12 of the time resolution presented 
above.  This new simulation, H277, is a Milky-Way like galaxy with a rather 
quiet merger history and a total halo mass of 7.1$\times$10$^{11}$ \Msun 
~(see Table~\ref{simsum}) at $z$ = 0.  For the same step interval as was 
used above, the gas accretion results for this galaxy agree well with the 
trends shown for the previous four galaxies.  From Table~\ref{accr} it can 
be seen that the fraction of shock accreted gas lies nicely between DWF1 
and GAL1, as might be expected for this mass of galaxy.  

A change in time resolution 
has an effect at two different stages of our analysis.  First, as just 
mentioned, we may miss particles with short cooling times that should have 
been counted by our ``shocked'' definition.  Second, ``clumpy'' gas
is identified as belonging at any output time step to a halo other than 
the main galaxy halo.  We are likely to miss gas particles that belong to 
another galaxy for less than $\sim$320 Myr (the time between output steps),
e.g., because they are subsequently lost from the galaxy due to SNe.  The 
result of missing this clumpy accretion will be to increase the number 
of smoothly accreted particles instead. 

These two stages of our analysis are not independent of one another, 
since the sample of shocked gas is drawn only from those gas particles 
that are identified as smooth accretion.  We wish to quantify them 
separately.  We first 
test how the number of particles identified as ``clumpy'' changes as a 
function of time resolution.  We investigate the change at 12x the original 
step interval used above (i.e., fine sampling corresponds to $\sim$27 
Myr resolution per output).  
 
When using our standard step interval, the smooth gas makes up 
73\% of all the gas accreted by galaxy H277 over the history of the 
trace.  By increasing the output interval resolution by a factor of 12, 
the gas identified as smoothly accreted declines, so that it composes 
66\% of all the gas accretion over the galaxy's history, for a total 
decline of 7\%. 
Thus, by definition, the fraction of gas accreted as ``clumpy'' over 
the galaxy's history must increase by 7\%.  

This result means that our absolute numbers stated above for 
the fraction of gas that is smoothly accreted to each galaxy is an 
upper limit.  An examination of the accretion history of this galaxy 
shows smooth accretion with characteristic spikes indicative of 
mergers.  The amount of gas in these spikes identified as ``smooth'' 
decreases in our fine time step limit, but the spikes still remain, 
suggesting we are still missing some gas in mergers.  By generously 
estimating the mass of gas contributing to these accretion spikes, 
we find that we are missing only an additional 1\% of clumpy gas 
accretion.  Thus, we agree with previous studies \citep[K05,][]
{Murali02} that find that the majority of gas mass 
is accreted smoothly rather than in mergers.  

Next, we use the ``smooth'' particles identified in the standard 
output resolution
(320 Myr outputs, where smooth gas accounts for 73\% of all the 
gas accreted over the history of the galaxy) to isolate the effect of 
output interval on the fraction of particles identified as shocked.  
At finer step intervals, the fraction of smooth gas that 
is shocked increases by 2.3\%.  This increase is quite small, and 
occurs primarily at $z$ $>$ 2, likely to due to the fact that we do not 
adequately sample cooling times at high redshift.  
To verify this, we calculated the instantaneous cooling rates for particles 
in an annulus between 30-40 kpc (as close to the disk as we trace them) 
and inside 10 kpc of R$_{vir}$ at each of our ``standard'' analysis  
intervals.  For H277, the virial radius does not reach 30 kpc until 
$z$ = 2.65, and the cooling times of the particles within R$_{vir}$ are 
lower than our standard step resolution of 320 Myr until this redshift.  
After this redshift, however, the galaxy grows quickly and the cooling 
times are longer than our output intervals, so that we are capable of 
easily identifying shocked gas.  
Thus, it is not surprising that the addition of shocked gas when 
increasing time resolution comes at high redshifts.  

We have done a similar cooling time analysis for our four main galaxies.  
In the case of DWF1, MW1, and GAL1, we do not fully resolve cooling 
times within the halos until redshifts 3, 3, and 4.5, respectively.  
However, because DWF1 is not massive enough to support a stable shock 
until $z$ $\sim$0.6, our time step undersampling at high $z$ has no 
impact on our resulting fraction of 
shocked gas.  For MW1 and GAL1, we can expect about a 5\% change, as 
we found in h277, and perhaps less since we resolve their 
cooling times to higher redshift than in the H277 case.  

As an interesting note, our lowest mass galaxy, H579, always has 
cooling times longer than our standard output step interval (though 
it is never massive enough to shock, so this is not important for 
our results).  It may seem odd at first that this galaxy has such 
long cooling times, but it is a result of the lower densities in 
this halo.  Our UV background is capable of ionizing much of the 
hydrogen gas in the halo of this galaxy, causing it to be 
inefficient at cooling.  We verified that turning off our 
UV background lowers the instantaneous cooling times substantially, 
by almost an order of magnitude \citep{efstathiou92, Quinn96}.

\subsection{Entropy Generation}
\label{entropy}

By definition, a shock is a phenomenon that leads to a sharp discontinuity 
in the physical properties of a gas.  Due to the nature of smoothed particle 
hydrodynamic codes, these properties for any given gas particle will be a 
smoothed estimate based on the properties of its nearest neighbors.  Thus, 
there is some concern that SPH will do a poor job of resolving shocks.  

{\sc Gasoline} adopts the standard SPH artificial viscosity formulation 
\citep{monaghan92}, with a correction to detect and reduce viscous transport 
in shearing regions \citep{balsara95}. 
\citet{gasoline} investigated {\sc Gasoline}'s performance in this area 
with both the standard \citet{sod} shock tube test, and a spherical 
adiabatic collapse test \citep{evrard88}.  It was found that {\sc Gasoline} 
does an excellent job of matching the analytic solutions to within a few 
particle separations, and matches the post-shock entropy values very well.  

It can be possible for low numerical resolution to artificially broaden 
shocks (due to entropy generated by the artificial viscosity used in SPH), 
allowing gas to radiate energy while passing through a shock rather than 
heating and undergoing an entropy jump \citep{ht00}.  This broadening is 
unlikely to be an issue at the high resolutions used here 
\citep{springel05}, but we verified numerical convergence by using a lower 
resolution version of our MW1 run, MW1.lo, and performing the same analysis.  
This run has 1/8 of the mass resolution and 1/2 the force resolution of MW1 
\citep[and is the same galaxy discussed in detail in][]{G07}.  
Table~\ref{accr} demonstrates that the shocked fraction in this run is in 
excellent agreement with MW1.  Since our shock definition includes a 
temperature criterion, this implies that we are not missing high temperature 
gas, and thus that we have a sufficient numerical resolution to identify 
entropy jumps and are not missing heating in shocks.  Hence, our cold flow 
gas accretion is not an artifact of numerical resolution.

\subsection{An Alternate Shock Definition}
\label{temp}

Previous groups who have attempted to investigate the role of cold flows 
in galaxy gas accretion have used a different criterion to identify shocked 
and unshocked gas, adopting a strict temperature cutoff.
We examine in this section how our results differ if we use the same  
temperature definition as other authors.  

\begin{figure*}
\plotone{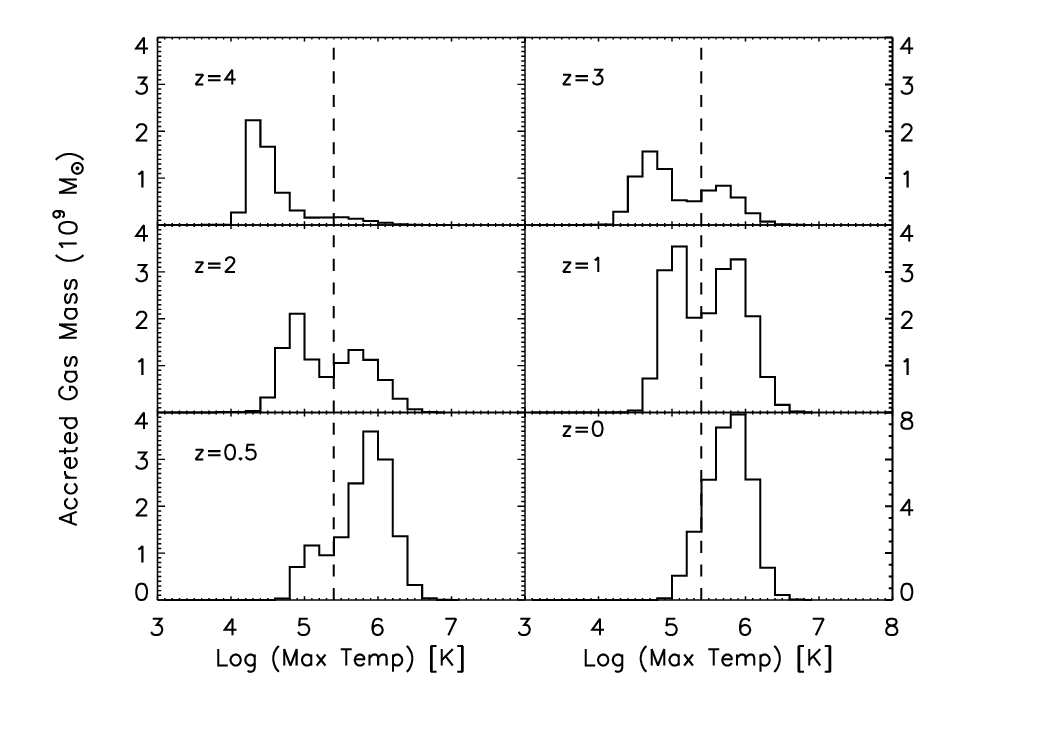}
\caption{ This figure shows histograms of all of the {\it smoothly} 
accreted gas between successive redshift intervals for our MW1 galaxy.  
The dashed vertical line shows the temperature divide used to separate 
hot and cold accretion modes used by K05 and OPT08. }
\label{mwhist}
\end{figure*}

K05 and OPT08 have noted that 
the temperature distribution of accreted particles can be bimodal, with 
a minimum at 2.5$\times$10$^5$ K.  They use this minimum to separate gas 
into cold accretion and hot accretion modes.  We also see 
this minimum in the temperature distribution, as shown in Fig.~\ref{mwhist} 
for our MW1 galaxy.  This figure shows the maximum temperature of 
gas particles that have been accreted to the galaxy between successive 
redshift intervals.  Up until $z$ = 0.5, there is a strong bimodality. 
The dashed vertical line shows the separation at 2.5$\times$10$^5$ K used 
by K05, K08, and OPT08.  Our accreted gas particles clearly demonstrate this 
same minimum.  These two groups have previously noted that this minimum 
is a function of the cooling curve, and has therefore has little direct 
relation to the actual modes of hot or cold accretion.  However, it is 
a convenient point at which to separate the gas into hot or cold 
accretion modes.  

To assess how our results are affected by these definitions, we repeat 
the above analysis to identify shocked gas in our smoothly accreted gas 
component, using the temperature criterion of K05 and OPT08.  
The same set of smoothly accreted particles is used, so that the only change 
between the two definitions will be in the resulting shocked or unshocked 
fraction (or, more appropriately, hot and cold fraction).  The 
difference resulting from these two separate definitions is remarkably 
small.  The temperature criterion leads to slightly more hot gas being 
accreted to the galaxy than shocked gas, but the largest difference is 
only 7\%, in the case of GAL1.  The remaining galaxies have differences 
of 3\% or less.  H579 never reaches the temperature cutoff identified 
by K05, and thus both studies agree than it has never accreted 
shocked/hot gas.  

These tests lead us to conclude that the two different definitions 
yield very similar results.  This agreement may be partly due to the 
fact that we also impose some sort of temperature criterion (requiring 
that our gas not only undergo an entropy increase, but reach 3/8 
T$_{vir}$ as well).  However, this result was not guaranteed.  We 
emphasize that the temperature criterion alone is based on the physics 
of the cooling curve, and does not reflect the physics of the gas 
accretion shocks.  Hence, we choose to add an entropy 
criterion to our definition, as it is unique to each galaxy halo and 
the physics of the various accretion modes.

\subsection{The Role of the UV Background}
\label{uv}

\citet{G07} found that the mass function of our galaxies is resolved down 
to a minimum of 64 DM particles.  For the runs used here, that corresponds 
to total halo masses of 8.1$\times$10$^6$ \Msun, 6.1$\times$10$^6$ \Msun, 
4.9$\times$10$^7$ \Msun, and 1.9$\times$10$^8$ \Msun ~for H579, DWF1, MW1, 
and GAL1, respectively.  These 
masses are well below the limit where the UV field will suppress galaxy 
formation \citep{G07,efstathiou92,Quinn96,thoul,gnedin00,hoeft06},   
and thus we avoid issues due to sub-resolution merging.  That is, there 
should be no unresolved halos that are being counted incorrectly as 
``smooth'' accretion.  This assertion was verified by running the same 
analysis on a lower resolution version of our MW1 galaxy, MW1.lo, with 
a mass resolution of 1.9$\times$10$^8$ \Msun ~at 64 DM particles.  This 
mass is just below the limit that the UV field should suppress galaxy 
formation.  Indeed, the smooth gas fraction agrees to within 
a few percent with MW1 (see Table~\ref{accr}).  MW1.lo does have a five 
percent higher fraction of smooth accretion, but given fluctuations in 
the simulations, this is excellent agreement. 

The dominant effect of the UV background in our simulations is to suppress 
galaxy formation at low masses, and hence lower the amount of clumpy 
accretion.  Gas which would 
have otherwise been in small halos is now free to be smoothly accreted 
to our galaxies.  
This was confirmed by running MW1.lo with the UV background off (and 
cooling and SF turned off as well).  The results for this galaxy, MW1.ad, 
are shown in Table~\ref{accr}.  It is clear from the first two columns 
that the amount of ``clumpy'' accretion material increases from about 
20\% to about 50\%.  This confirms that our UV background is partly 
responsible for reducing the amount of gas accreted from other halos,
and hence increasing the amount of gas available for smooth accretion.

The more interesting effect of the UV background is on the amount of 
shocked gas accretion to the galaxy.  Our UV background (in combination 
with the mean baryonic density of the universe as a function of redshift) 
sets a temperature floor in the IGM, and hence an 
entropy floor for IGM gas.  For a galaxy to be capable of shocking accreted 
gas, the virial entropy of the galaxy halo must be above this background 
entropy by a certain factor.  Hence, based on our definitions, the UV 
background sets a minimum halo mass that will be able to shock, as function 
of time.  
 
\subsection{Cooling}
\subsubsection{The Adiabatic Case}
\label{adiabatic}

\citet{BD03} and DB06 showed that the shock radius in the case of adiabatic 
accretion (i.e., no cooling) is well matched to the virial radius in a
spherically symmetric case. 
They also demonstrated that when cooling is turned on, the same galaxy 
is not initially massive enough to support a stable shock, so that accreted 
gas shocks at the disk of the assembling galaxy.  Only after a critical 
mass is reached can the galaxy support a shock.  

We have again used MW1.ad with the UV background, cooling, and SF turned 
off (i.e., the adiabatic case) to test these results.  As noted above, the 
amount of smooth accretion drops for MW1.ad, as much of the gas is able 
to remain bound to its dark matter halo.  However, from Table~\ref{accr} 
(last two columns) it is clear that the amount of smooth gas accretion that 
is shocked also increases in this case, increasing to 84\%.  This behavior 
is expected, and is similar to the standard spherical model in which all 
gas gets heated to the virial temperature of the galaxy as it is accreted.  
The fact that there is unshocked gas at all is due to the fact that this 
simulation is not spherically symmetric, and filaments of cold gas still 
exist that are capable of penetrating well inside of the halo without 
shocking.

\begin{figure}
\plotone{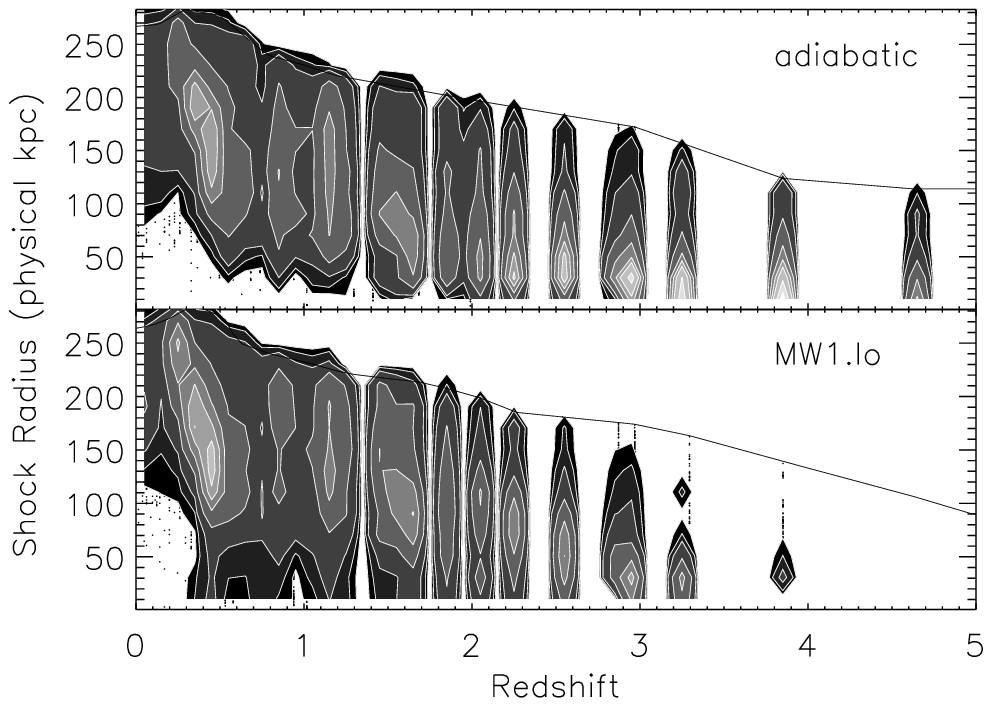}
\caption{These contours show the radii as which particles are identified 
as shocked, for our MW1.lo and adiabatic, MW1.ad, runs.  Black represents the 
lowest concentration of particles at a given time, and white peaks are 
the highest.  The solid lines show the virial radius as a function of time.}
\label{shockcomp}
\end{figure}

Fig.~\ref{shockcomp} shows the radii at which particles are shocked inside 
of R$_{vir}$, as a function of time, for the MW1.lo and adiabatic, MW1.ad, 
runs.  Contours are used for clarity, with black representing the least 
dense concentration of particles, and white peaks the most dense.  The 
solid lines show the virial radii of the galaxies as a function of time.  
The main difference between the two runs is at high redshift.  In the 
adiabatic case, particles are able to shock inside the virial radius even 
at high $z$, while in the MW1.lo run there is no shocking at the highest 
redshifts.  Shocks begin to occur near the disk at redshift 4, and shocked 
particles fill the space between the disk and the virial radius at 
lower redshifts.  The radius with the highest concentration of shocked 
particles moves outward with time in both cases.

\subsubsection{Metal Line Cooling}
\label{mc}

Metal line cooling has recently been added to {\sc Gasoline} (Shen et al., 
in prep.).  MW1.lo was rerun with metal cooling on, to find if the new 
cooling has any effect on the results here.  However, the smoothly accreted 
gas (be it shocked or unshocked) has nearly primordial metallicities, and 
the resulting fractions of shocked and unshocked accreted gas are identical 
to the results for MW1.lo with only H and He cooling.  Likewise, the 
fraction of disk stars formed from shocked and unshocked gas is also 
identical to MW1.lo.

\section{Discussion \& Conclusions} 
\label{conclusions}

We have tracked the history of gas accreted to four high resolution 
disk galaxies that span two orders of magnitude in galaxy mass. 
For all of these galaxies, early disk growth can be predominantly 
attributed to cold gas, and up to Milky Way masses the SF of the stellar 
disk is dominated by unshocked, cold flow gas accretion at essentially 
all times.  This dominance of cold gas in the formation of disk stars 
can exist even when the the majority of gas accreting to the virial 
radius is being shocked and reaching high temperatures.  At galaxy masses 
above L$^*$, an early phase of disk growth prior to $z$ = 1 can be 
attributed to the presence of cold gas accretion.

We have examined gas temperature histories in relation to the virial 
temperature of the halo that they are accreted to.  While the shocked 
component is capable of reaching temperatures near the virial 
temperature of the halo, the unshocked component of gas accretion 
remains below 3/8 of the virial temperature at all times.  The 
important role of unshocked, cold flow gas in the growth of galaxies 
is consistent with previous studies that adopt an explicit temperature 
cutoff, determined by the cooling curve, to identify cold flow gas.  
We confirm that the ability of smoothly accreted gas to shock is a strong 
function of mass.  However, our focus here has been to study the 
subsequent role of this cold gas accretion in the growth of stellar 
disks.  Because the unshocked gas does not reach the same high 
temperatures as shocked gas, it is capable of cooling to the disk 
faster and forming stars.  

It is important to note that this early cooling of unshocked gas 
to the disk is 
not a manifestation of the historic ``overcooling'' problem.  This 
can be seen by examining the results from our stellar mass - 
metallicity relationship, which matches the observed relation at 
both low and high redshifts \citep{Brooks07,maiolino08}.  As discussed 
by both \citet{Brooks07} and \citet{maiolino08}, this match is achieved 
by overcoming the overcooling problem with both high resolution and a 
physically motivated SN feedback model.  Previous works consistently 
produced metallicities too high when compared to observations at high 
$z$, due to overcooling and a rapid consumption of gas and subsequent 
metal enrichment.  Instead, our SN feedback scheme regulates the SFRs 
of our galaxies, preventing them from too quickly consuming their gas.
The early SF seen in our more massive galaxies is due instead to the 
accretion of cold gas that can cool rapidly to the disk to form 
stars.

Despite running our simulations in a fully cosmological context that 
includes galaxy mergers, the dominant gas supply is from smoothly 
accreted gas that has never belonged to another galaxy halo.  For 
a Milky Way mass galaxy with a quiescent merging history like the 
galaxies shown here, only $\sim$25\% of the accreted gas is acquired 
from other galaxy halos, resulting in a similar fraction of the disk 
stars forming from this gas.

It is our goal to assess the impact that cold flow gas accretion may 
have in altering current models of disk galaxy formation.  A direct 
comparison with semi-analytic models is difficult to do and beyond the 
scope of this paper.  Current SAMs include the possibility of rapid 
gas cooling for low mass galaxies.  
It has been suggested that this regime is similar to the cold flow 
gas accretion seen in simulations \citep{croton06b}.  
\citet{cattaneo07} performed a detailed comparison of the GalICS 
SAM \citep{galics} to the SPH results of K05, and found that the 
cold gas accretion rates in the two schemes were in agreement.  
This implies that the amount of cold gas currently available in SAMs 
may agree well with simulations, though this needs to be confirmed 
with further comparisons.  

However, the results in this paper may emphasize a separate problem 
within the cold flow regime.  Even in the case of rapid cooling in 
SAMs, the temperature of the gas is initially set to the virial 
temperature of the galaxy halo, and cooling times are based on this 
assumption.  
Prior to the development of a hot halo, the free fall time determines 
when the gas reaches the disk in either the SAMs or the simulations.  
However, SAMs do not yet model the impact of filaments on galaxy growth.  
Our two most massive galaxies, which are able to develop a hot halo
at high redshifts \citep[4 to 6; see also][]{derossi08}, are still 
capable of accreting cold gas that never approaches the virial 
temperature of the halo in filaments at high $z$.  
These filaments allow a significant amount of cold gas mass to 
accumulate inside of the galaxy, even for massive galaxies capable of 
supporting a shock at early times.  This cold gas cools quickly to the 
disk to form stars.  This rapid settling to the disk allows for the 
building of stellar disks at higher redshifts than predicted if all 
gas is heated to the virial temperature, as in the standard model.

We have shown that the discrepancy in temperatures between the 
shocked and unshocked gas components leads to significantly longer 
cooling times for the shocked gas (note that their densities will 
be discrepant as well, since most of the unshocked component are 
found in dense filaments, exacerbating the difference in cooling 
times).  If the unshocked gas had reached the virial temperature, 
the star formation rates in the disk would have been delayed (in 
the cases shown here, until after $z$ = 1).  A study of the 
angular momentum content of the accreted gas will be examined in 
future work, but the presence of cold flow gas accretion, 
particularly in filaments in massive halos, allows for the 
building of large, massive disks at high $z$, in agreement with 
observations.


\acknowledgments

We would like to thank A. Dekel, D. Keres, J. Dalcanton, S. White, and
Q. Guo, for helpful conversations during this project.  AB, FG, and TQ 
were supported by NSF ITR grant PHY-0205413.  AB acknowledges support 
from a WA Space Grant fellowship.  FG acknowledges support from a 
Theodore Dunham grant, HST GO-1125, NSF grant AST-0607819 and NASA 
ATP NNX08AG84G.
Simulations were run at TACC, SDSC, Cineca, and NAS.  


\end{document}